\documentclass[fleqn,usenatbib]{mnras}

\usepackage[T1]{fontenc}
\usepackage{ae,aecompl}

\usepackage{graphicx}    
\usepackage{amsmath}    
\usepackage{amssymb}    
\usepackage{stfloats}
\usepackage{pdflscape} 
\usepackage{eso-pic}

\AddToShipoutPictureBG*{%
  \AtPageUpperLeft{%
    \hspace{0.75\paperwidth}%
    \raisebox{-3.5\baselineskip}{%
      \makebox[0pt][l]{\textnormal{DES-2022-0693}}
}}}%

\AddToShipoutPictureBG*{%
  \AtPageUpperLeft{%
    \hspace{0.75\paperwidth}%
    \raisebox{-4.5\baselineskip}{%
      \makebox[0pt][l]{\textnormal{FERMILAB-PUB-22-604-PPD}}
}}}%

\title[Comparison of CNN Classifications]{Lessons Learned from the Two Largest Galaxy Morphological Classification Catalogues built by Convolutional Neural Networks}
\author[Ting-Yun Cheng et al.]{Ting-Yun Cheng$^{1}$,\thanks{E-mail:ting-yun.cheng@durham.ac.uk}
H.~Dom{\'\i}nguez S{\'a}nchez$^{2}$,
J.~Vega-Ferrero$^{3}$,
C.~J.~Conselice$^{4}$,
\newauthor
M.~Siudek$^{5,2}$,
A.~Arag\'on-Salamanca$^{6}$,
M.~Bernardi$^{7}$,
R.~Cooke$^{1}$,
L.~Ferreira$^{6}$,
\newauthor
M.~Huertas-Company$^{3,8}$,
J.~Krywult$^{9}$,
A.~Palmese$^{10}$,
A.~Pieres$^{11,12}$,
\newauthor
A.~A.~Plazas~Malag\'on$^{13}$,
A.~Carnero~Rosell$^{3,11,14}$,
D.~Gruen$^{15}$,
D.~Thomas$^{16}$,
\newauthor
D.~Bacon$^{16}$,
D.~Brooks$^{17}$,
D.~J.~James$^{18}$,
D.~L.~Hollowood$^{19}$,
D.~Friedel$^{20}$,
E.~Suchyta$^{21}$,
\newauthor
E.~Sanchez$^{22}$,
F.~Menanteau$^{20,23}$,
F.~Paz-Chinch\'{o}n$^{20,24}$,
G.~Gutierrez$^{25}$,
G.~Tarle$^{26}$,
\newauthor
I.~Sevilla-Noarbe$^{22}$,
I.~Ferrero$^{27}$,
J.~Annis$^{25}$,
J.~Frieman$^{25,28}$,
J.~Garc\'ia-Bellido$^{29}$,
\newauthor
J. Mena-Fern{\'a}ndez$^{22}$,
K.~Honscheid$^{30,31}$,
K.~Kuehn$^{32,33}$,
L.~N.~da Costa$^{11}$,
M.~Gatti$^{7}$,
\newauthor
M.~Raveri$^{7}$,
M.~E.~S.~Pereira$^{34}$,
M.~Rodriguez-Monroy$^{22}$,
M.~Smith$^{35}$,
\newauthor
M.~Carrasco~Kind$^{20,23}$,
M.~Aguena$^{11}$,
M.~E.~C.~Swanson,
N.~Weaverdyck$^{26,36}$,
P.~Doel$^{17}$,
\newauthor
R.~Miquel$^{37,5}$,
R.~L.~C.~Ogando$^{12}$,
R.~A.~Gruendl$^{20,23}$,
S.~Allam$^{25}$
S.~R.~Hinton$^{38}$,
\newauthor
S.~Dodelson$^{39,40}$,
S.~Bocquet$^{15}$,
S.~Desai$^{41}$,
S.~Everett$^{42}$,
V.~Scarpine$^{25}$
\\
\\
\parbox{\textwidth}{\centering \textsc{\Large  } \\ \centering \textit{Author affiliations are listed at the end of this paper} }
}
\date{Accepted XXX. Received YYY; in original form ZZZ}

\pubyear{2022}

\begin{document}
\label{firstpage}
\pagerange{\pageref{firstpage}--\pageref{lastpage}}
\maketitle

\begin{abstract}
We compare the two largest galaxy morphology catalogues, which separate early and late type galaxies at intermediate redshift. The two catalogues were built by applying supervised deep learning (convolutional neural networks, CNNs) to the Dark Energy Survey data down to a magnitude limit of $\sim$21 mag. The methodologies used for the construction of the catalogues include differences such as the cutout sizes, the labels used for training, and the input to the CNN - monochromatic images versus $gri$-band normalized images. In addition, one catalogue is trained using bright galaxies observed with DES ($i<18$), while the other is trained with bright galaxies ($r<17.5$) and `emulated' galaxies up to $r$-band magnitude $22.5$. Despite the different approaches, the agreement between the two catalogues is excellent up to $i<19$, demonstrating that  CNN predictions are reliable for samples at least one magnitude fainter than the training sample limit. It also shows that morphological classifications based on monochromatic images are comparable to those based on $gri$-band images, at least in the bright regime. At fainter magnitudes, $i>19$, the overall agreement is good ($\sim$95\%), but is mostly driven by the large spiral fraction in the two catalogues. In contrast, the agreement within the elliptical population is not as good, especially at faint magnitudes. By studying the mismatched cases we are able to identify lenticular galaxies (at least up to $i<19$), which are difficult to distinguish using standard classification approaches. The synergy of both catalogues provides an unique opportunity to select a population of unusual galaxies.

\end{abstract}


\section{Introduction}
\label{sec:intro}

Galaxy morphology describes the visual features of a galaxy and the structure of its light distribution.  Both of these properties are strongly connected with its formation history \citep[e.g.,][]{Holmberg1958, Dressler1980}. In addition, galaxy morphologies are intimately related to their stellar populations, likewise for galaxy stellar masses, star formation rates, ages, and metallicities \citep[e.g.,][]{conselice2006,Pozzetti2010,Wuyts2011,Huertas-Company2016}. Therefore, obtaining a large number of galaxies with robust morphological classifications is of great importance for understanding galaxy evolutionary history and stages.

Traditionally, the morphological classification of galaxies has been based on visual inspection \citep{deVaucouleurs1959, Sandage1961, deVaucouleurs1964, Fukugita2007, Nair2010, Baillard2011}. However, during the past decades, there was a significant increase in the sizes of galaxy datasets, from the Hubble Space Telescope, or the Sloan Digital Sky Survey, to current surveys such as Dark Energy Survey which includes hundreds of millions of galaxies \citep[][hereafter, DES]{DES2005, DES2016}.  The number of observed galaxies will increase even more with the advent of future surveys such as the Euclid Space Telescope \citep[]{laureijs2011} and the Vera Rubin Observatory Legacy Survey of Space and Time \citep{Ivezic2019}, making it impossible to  visually classify such extremely large datasets, even with Citizen Science tools such as Galaxy Zoo \citep{Lintott2008,Lintott2011,Willett2013}. Thankfully, we can now utilise machine learning techniques, which have been applied to a variety of astronomical studies with great success since the 1990s: e.g., star-galaxy separation \citep[][]{Odewahn1992, Weir1995,Soumagnac2015}, photo-z estimation \citep[][]{Collister2004,Alarcon2021,Soo2021,Schuldt2021}, galaxy structural measurements \citep{Tohill2021}, strong lensing identification \citep[][]{Jacobs2017, Petrillo2017, Lanusse2018, Cheng2020b}, finding galaxy mergers \citep[][]{Bottrell2019, Ferreira2020}, and morphological classification of galaxies \citep[][]{Lahav1995, Banerji2010,Huertas-Company2018, Dominguez-Sanchez2018, Siudek2018a,Siudek2018b, Walmsley2020, Cheng2020a, Hausen2020, Ghosh2020, Cheng2021-uml, Turner2021,Gupta2022}.

The main advantages of machine learning techniques, and in particular, of CNN algorithms, compared to light-profile parametric fitting (e.g., \citealt{Tarsitano2018, Everett2022}) are  the computation time, orders of magnitude faster once the modes have been trained, and the lack of a model assumption to describe the light profile (of particular importance for asymmetric galaxies, which are more frequent at higher redshift).

Machine learning techniques are applied to astronomical studies using a diversity of approaches, but there are not many comparisons of these different techniques to assess their successes and failures. In this work, we compare the results of the two morphological catalogues presented in \citet[][hereafter V21]{Vega-Ferrero2021} and \citet[][hereafter C21]{Cheng2021-cata}, constructed by applying CNNs to the DES data. The differences between the two approaches provide an excellent opportunity to assess, with great statistics, the impact of different deep learning methodologies (e.g., training labels, depth of the training sample, the inclusion of `emulated' images in the training sample, etc). The catalogues are the two largest morphological classification catalogues to date including $\sim27$ million (V21) and $\sim21$ (C21) million galaxies, respectively. The overlapping sample includes over 17 million galaxies. This comparison not only potentially further validates the classification of the two catalogues, but also provides a detailed analysis of different approaches within CNN galaxy morphology studies. Additionally, by combining the two catalogs one can get more than 30 million galaxies with morphological classification.

Since both catalogues have their own related papers, in this work we focus on their comparison. The arrangement of this paper is as follows. In Section~\ref{sec:data_catalogues}, we discuss the differences between the two catalogues in sample selection and methodology. 
We present a statistical comparison of the two catalogues in Section~\ref{sec:comp_statistics}. We discuss the agreement and disagreement between the two catalogues  in Section~\ref{sec:examination}. This part of the discussion is divided into bright and faint galaxies and a detailed analysis of the mismatched cases. Finally, a comparison of the physical properties and structural measurements  of the two classes reported by each catalogue is discussed in Section~\ref{sec:comp_physics}. We summarise our results in Section~\ref{sec:summary}.

\section{Data \& Catalogues}
\label{sec:data_catalogues}

In this work, we compare the two largest galaxy morphological classification catalogues to date: V21 which contains $\sim$27 million galaxies with $r$-band magnitude brighter than 21.5, and C21 which includes $\sim$21 million galaxies with an $i$-band magnitude range $16\le{i}<21$ and at redshift $z<1$. Both catalogues are built by analyzing DES data, which is a wide-field optical imaging survey covering 5000 square degrees \citep[$\sim$1/8 sky;][]{Neilsen2019}. The coadd images have a spatial resolution of 0.263 arcsec per pixel and were taken by the Dark Energy Camera \citep{Flaugher2015}, which has a high quantum efficiency in the red wavebands ($>90$\% from $\sim$650 to $\sim$900 nm), and includes $\sim$ 300 million galaxies with a mean median depth of $g=24.33$, $r=24.08$, $i=23.44$ at a single-to-noise ratio S/N=10 \citep{Abbott2018}. The imaging data used in both works is DES Year 3 (Y3) data, but with different selection criteria for the initial samples (see Section~\ref{sec:data_selection}) from the DES Y3 GOLD catalogue \citep[][]{Sevilla-Noarbe2021}.

If not specified, physical properties of galaxies, such as apparent magnitude, redshift, colour, etc., shown in this work are from the DES Y3 GOLD catalogue. Additional stellar mass information is obtained by running the \texttt{LePhare} code \citep{Arnouts2011} using \citet{Bruzual2003} templates, three different metallicities (including solar), Chabrier Initial Mass Function (IMF), and exponentially declining star formation histories \citep[similar to][]{Palmese2020}\footnote{\label{Palmese_Cata}This unpublished catalogue is built by A. Palmese}.

Another catalogue built by \citet[][hereafter, T18]{Tarsitano2018} provides structural measurements, such as S\'ersic index, ellipticity, etc. While writing this manuscript, a new parametric light-profile fitting model for DES Y3 GOLD was released \citep{Everett2022} which could serve as an additional comparison to the
predictions of the CNN. However, there are some assumptions in the derivation of those parameters\footnote{For example, the assumption of a DeVauculers profile or a prior of 0 for the \texttt{FRACDEV} parameter at the faint end.} which may need further examination. Therefore, in this work we limit the comparison of structural parameters with those from T18. In Section~\ref{sec:faint_galaxies}, we also use the measurements from the VIMOS Public Extragalactic Redshift Survey \citep[VIPERS;][]{Moutard2016a,Moutard2016b,Krywult2017,Scodeggio2018,Siudek2018b} to check the robustness of the morphological classifications.


\subsection{Data Selection}
\label{sec:data_selection}
\begin{figure*}{}
\begin{center}
\graphicspath{}
	\includegraphics[width=2\columnwidth]{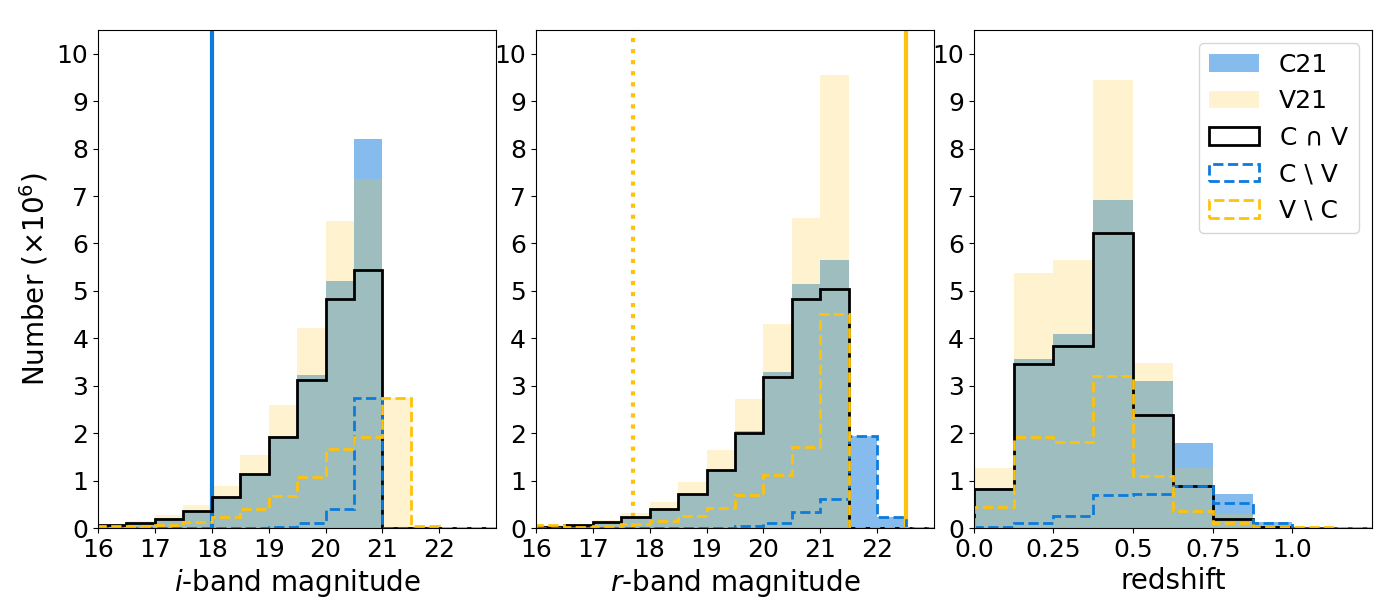}
   	\caption{Magnitude and redshift distributions of C21 sample (blue filled histogram), V21 sample (light yellow filled histogram), and different sub-sets of them. The overlapped area is represented as the mix of blue and yellow, i.e. slightly darker blue/green. The solid black lines show the intersection of the two catalogues. The blue dashed lines and the yellow dashed lines are for samples only in C21 (C$\backslash$V) and only in V21 (V$\backslash$C), respectively. The vertical lines indicate the upper magnitude limit of the training samples such that the blue solid line is for the one of C21 ($i<18$), the yellow dotted line is the magnitude limit of the  bright training sample in V21 ($r<17.7$), and the yellow solid line represents the limit of the final training samples in V21, including emulated faint galaxies ($r<22.5$).}
    \label{fig:comp_mag_z_hist}
\end{center}
\end{figure*}
Different selection criteria for constructing the  morphologically classified samples were applied in V21 and C21. The first significant difference are the filters used to define the upper limit of the samples' brightness: V21 imposed $r<21.5$ while in C21 the selection was $16\le{i}<21$. Fig.~\ref{fig:comp_mag_z_hist} shows the magnitude and redshift distribution of the two catalogues. The observed magnitudes are from the DES Y3 GOLD catalogue, measured in an elliptical aperture shaped by the Kron radius in the $i$-band and the $r$-band.  The photometric redshifts  are obtained by the Directional Neighbourhood Fitting (DNF) algorithm \citep{DeVicente2016}. Galaxies in C21 with an $i$-band magnitude brighter than $18$ (the limit for their training sample, see section \ref{sec:method}) contribute to 3.5\% of the total sample. In V21, by including `emulated' faint galaxies (see (3) of Section~\ref{sec:method}), the magnitude limit of their training samples is $r<22.5$, covering the full magnitude range of the two catalogues. Fig.~\ref{fig:comp_mag_z_hist} shows that the use of different filters for selecting the initial samples results in $\sim3$ million and $\sim2$ million unique galaxies in V21 and C21, respectively (i.e., included in V21 and not in C21, and vice-versa). Additionally, C21 applied a redshift cut at $z<1$, because their classifier, which is trained with bright galaxies at low redshift, has difficulty discriminating two morphology types at $z\ge1$ (see C21, Section 5.3). This difference results in $\sim$43,000 galaxies included in V21 but not in C21.

Second, the two works applied several flags from the DES Y3 GOLD catalogues, which were not identical. For example, {\texttt{EXTENDED\_CLASS\_COADD}} distinguishes point-like objects and extended objects. V21 selected a sample with {\texttt{EXTENDED\_CLASS\_COADD}} $>1$ including medium and high confidence galaxies, while in C21, only high confidence galaxies were chosen ({\texttt{EXTENDED\_CLASS\_COADD}} $=3$). The different criteria in several flags result in significantly more initial samples in V21 than C21.

Finally, a cut in half-light radius  ($>2.8$ pixels) was applied in V21 to avoid giving a morphological classification to galaxies with not enough spatial resolution. This removes $\sim$ 1.36 million galaxies from V21 which are included in C21 (corresponding to $\sim$6\% of the total samples from C21).

Overall, the different selection criteria applied results in an overlap of $\sim$17 million galaxies, and $\sim$ 9 million and $\sim$ 3 million unique galaxies in V21 and C21, respectively. The union of the two catalogues increases the total number of morphological classifications to $\sim$30 million galaxies.

\subsection{Methodology}
\label{sec:method}
Apart from the sample selection and the CNN architectures, there are also significant differences in the methodology, including: (1) cutout sizes; (2) training labels; (3) brightness of the training sample; (4) the input to the CNN.

(1) \textbf{Cutout sizes:} V21 applied variable cutout sizes which are $\sim$11.4 times the half-light radius  centred on the target galaxy. Then, the images are re-sampled (down or up-sampled, depending on the size of the resulting cutout) into a size of 64$\times$64$\times$3 (where the 3 last channels correspond to the $g$, $r$ and $i$ bands). On the other hand, C21 estimates the size of a galaxy using their own algorithm\footnote{The size was estimated by counting the pixels that have signal-to-noise $>1\sigma$.}. Galaxies with a size smaller than or equal to 30$\times$30 pixels, a direct cutout with a size of 50$\times$50 pixels is made. For larger galaxies, a cutout of size 200$\times$200 pixels is made and then re-sampled into a size of 50$\times$50 pixels.

(2) \textbf{Training labels:} Since there were no morphological classifications for DES galaxies with which to train the deep learning models,  both studies use the morphological classifications from Sloan Digital Sky Survey (SDSS) imaging. In particular, V21 used the T-Type\footnote{T-Type is a classification scheme that uses continuous values (from -6 to $\sim$10) to categorise galaxy morphology. Elliptical galaxies have negative T-Types and spiral galaxies positive values.} presented in \citet[][hereafter, DS18]{Dominguez-Sanchez2018}, which are based on a deep learning model trained on the T-Type provided by the visual classification of \citet{Nair2010}.  The V21 training sample  was labeled as early-type galaxies (ETG; for galaxies with T-Type$<-0.5$) or late-type galaxies (LTG; for galaxies with T-Type$>0.5$). Galaxies with intermediate T-Types ($-0.5<$T-Type$<0.5$), where the scatter in the T-Type is large, were excluded from the training sample. This methodology helped the models to converge by removing galaxies with uncertain classifications. Note that by doing so, potential lenticular galaxies  are excluded in the training samples. On the other hand, C21 used the classifications of spiral and elliptical galaxies from Galaxy Zoo 1 catalogue \citep[GZ1;][]{Lintott2008,Lintott2011}, and therefore C21 separates elliptical (Es) from spiral galaxies (Sp). A correction in the GZ1 visual classifications was applied due to the better resolution and deeper images of the DES data compared to SDSS \citep[][also see \citealt{Fischer2019,DS2022}]{Cheng2020a}. This correction provides more accurate morphological labels for $\sim2.5\%$ of the training set used in C21, and excludes $\sim0.56\%$ of galaxies that are ambiguous for binary classifications. After this, the machine classifier in C21 is more sensitive to disk structures and proves to correctly identify disky galaxies with rounder and blurred features at faint magnitudes, which humans often incorrectly classify as elliptical galaxies (see the comparison with visual classification in Section 5.2 of C21). Note that, after this labelling correction, the Sp class in C21 includes galaxies with disk structures such as lenticular galaxies.

(3) \textbf{Brightness of the training sample:} Both catalogues use as the basis of their training sample bright SDSS galaxies ($r<17.7$ in V21 and $i<18$ in C21) with previous morphological classifications. While C21 only used the DES $i$-band images of bright galaxies as training sample, V21 `emulated' images at higher redshift. They included these faint galaxies in their  training sample keeping their original  morphological labels, i.e., without changing their `ground truth' despite their final appearance. Details on the emulation procedure are described in section 2.3 of V21, but in short, the original DES galaxy images were deconvolved by the PSF, flux and size corrections due to cosmological dimming were applied, as well as adding k-correction and evolutionary effects. Finally, noise was added and the images were re-convolved with their PSF. In V21 the models were tested on the `emulated' images (for which true labels were available) with an excellent performance (accuracy $\sim$ 97$\%$ up to $r<21.5$ mag), demonstrating that their CNN was able to predict correct morphological labels and detect features hidden to the human eye. Additional checks on the morphological classification on real DES faint galaxies were done using available data that correlates, to some degree, with morphology (see Section 5 in V21 and Section~\ref{sec:faint_galaxies} in this work.)

(4) \textbf{Input to the CNN:} V21 used $g,r,i$ band images (after normalising each band individually for each galaxy, to prevent the leak of colour information)  while C21 used only $i$-band images, but combined linear, logarithmic, and gradient images. This means that the V21 machine focuses on different structures that are shown in different wavelengths, while the C21 machine considers different structures emphasised in different scales, but uses a single band image.

\subsection{Classification Definition}
\label{sec:certain_type}
\begin{figure}{}
\begin{center}
\graphicspath{}
	\includegraphics[width=\columnwidth]{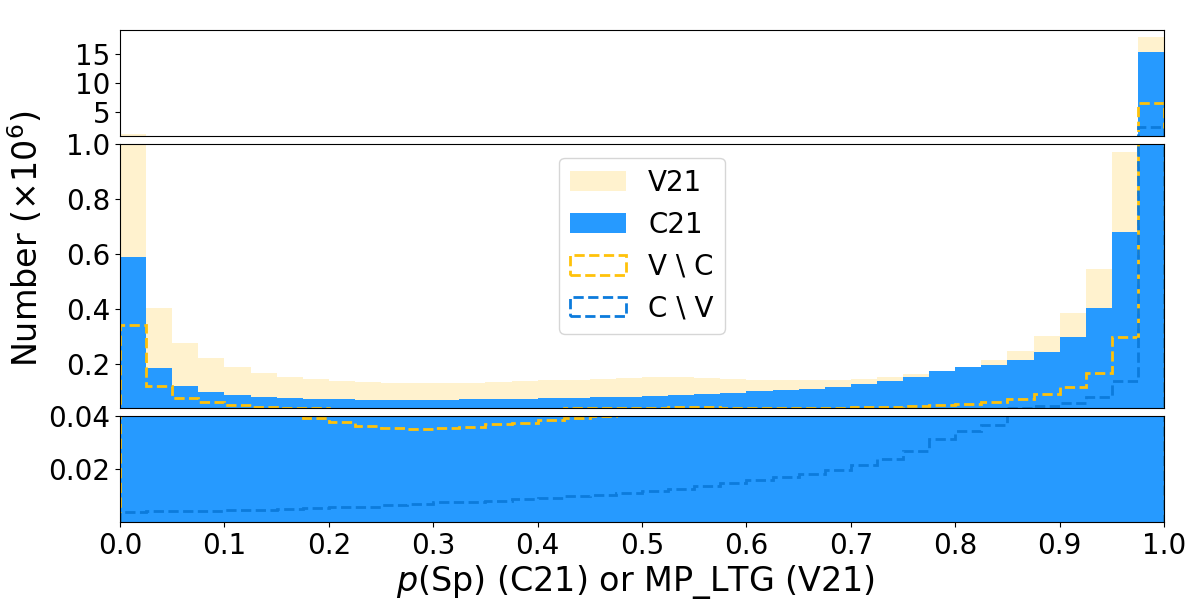}
   	\caption{Distributions of mean/median probabilities of being Sp (C21; $p$(Sp)) or LTG (V21; MP\_LTG= median($P_i$), where $P_i$ is defined in Section~\ref{sec:certain_type}). The light yellow and blue histograms represent all the galaxies in V21 and C21, respectively, while the dashed lines show the galaxies included in only one of the two catalogues (light yellow for V21 and blue for C21). The $y$-axis is truncated at different scales for optimal visualization.}
    \label{fig:probability}
\end{center}
\end{figure}

\begin{table*}
	\centering
	\begin{tabular}{lllllll}
		\hline
		\multicolumn{1}{l}{Number} & {V21} & {C21$^{b}$} & {C$\cap$V} & {C$\backslash$V} & {V$\backslash$C} & {C$\cup$V} \\
		\hline\hline
		\multicolumn{1}{l}{Total} & {26,971,945} & {21,119,107}  & {17,821,250} & {3,297,857} & {9,150,695} & {30,269,802} \\
		\multicolumn{1}{l}{Sp/LTG} & {19,789,809 (0.73)} & {17,532,564 (0.83)} & {11,830,693 (0.66)} & {2,938,910 (0.89)} & {7,100,049 (0.78)} & {21,869,652 (0.72)} \\
		\multicolumn{1}{l}{Es/ETG} & {2,332,097 (0.086)} & {1,309,229 (0.062)} & {689,332 (0.039)} & {35,999 (0.011)} & {649,727 (0.071)} & {1,375,058 (0.045)} \\
		\multicolumn{1}{l}{Uncertain} & {4,850,039 (0.18)$^{a}$} & {2,277,314 (0.11)} & {5,301,225 (0.30)$^{c}$} & {322,948 (0.098)} & {1,400,919 (0.15)} & {7,025,092 (0.23)$^{c}$} \\
		\hline
	\end{tabular}
	\caption{Number of total galaxies and certain morphological classes in each catalogue (as defined in Section~\ref{sec:certain_type}). Fractions are given in brackets. The `C$\cap$V' reports the number of galaxies in the intersection of the two catalogues; hence, the row of `Sp/LTG' shows the matched case of Sp \& LTG while the row of `Es/ETG' is Es \& ETG. The `C$\backslash$V' reports the galaxies included in C21 only, while `V$\backslash$C' shows the opposite case. Finally, `C$\cup$V' represents the union of the two catalogues. In this case, the row of `Sp/LTG' shows the sum of Sp, LTG, and Sp \& LTG while the row of `Es/ETG' is the sum of Es, ETG, and Es \& ETG.
	\\ \scriptsize $^a$ This number includes secure intermediate classification from V21. By excluding them, the number of galaxies with  uncertain classifications is 3,637,301 (0.13).
	\\ \scriptsize $^b$ When selecting samples with a confidence level > 2 in C21 (see Section 5.3 of C21), there are 13,312,568 galaxies (0.63 of the total number of galaxies). Within these selected samples, 10,259,513 (0.77 of the total selected samples) galaxies are Sp and 1,197,604 (0.090) galaxies are Es.
	\\ \scriptsize $^c$ Galaxies in this set are not assigned a certain classification in one of the two catalogues, or the classifications from the two catalogues disagree with each other.}
	\label{tab:statistics}
\end{table*}
\begin{figure*}{}
\begin{center}
\graphicspath{}
	\includegraphics[width=2\columnwidth]{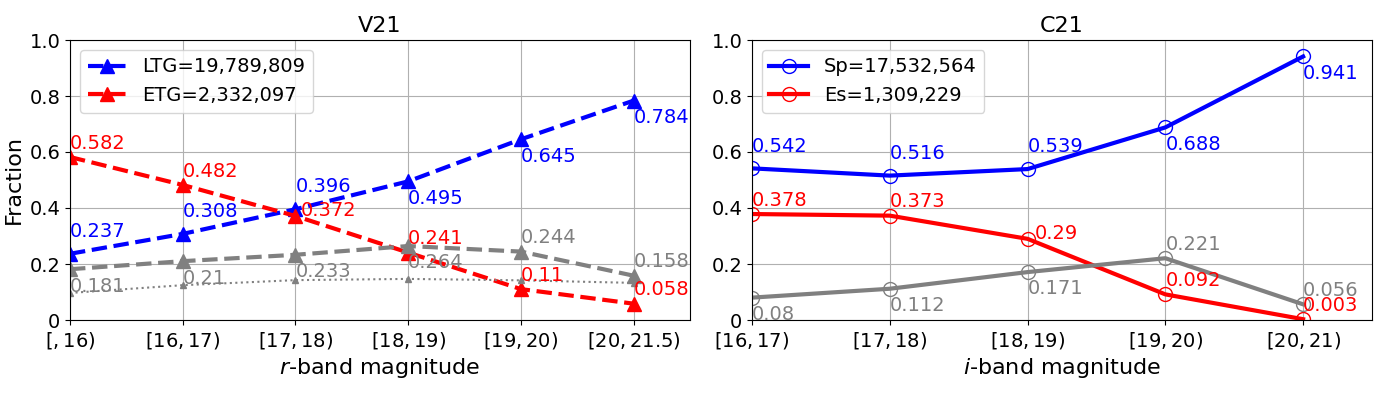}
   	\caption{Fraction of Es/ETG (red), Sp/LTG (blue) and galaxies with uncertain classifications (grey) with respect to the total in  $r$ and $i$-band magnitude bins for the V21 (left) and C21 (right) catalogues. The grey dotted line in the left panel (V21) are uncertain galaxies excluding `secure intermediate' classification (see Section~\ref{sec:certain_type} for the definition of each class).}
    \label{fig:frac_type_vs_mag}
\end{center}
\end{figure*}


\begin{figure*}{}
\begin{center}
\graphicspath{}
	\includegraphics[width=\columnwidth]{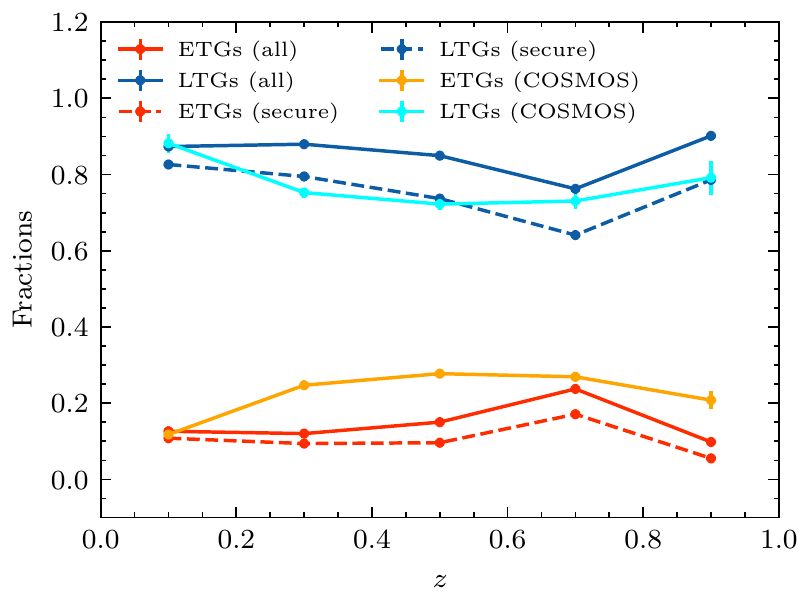}
	\includegraphics[width=\columnwidth]{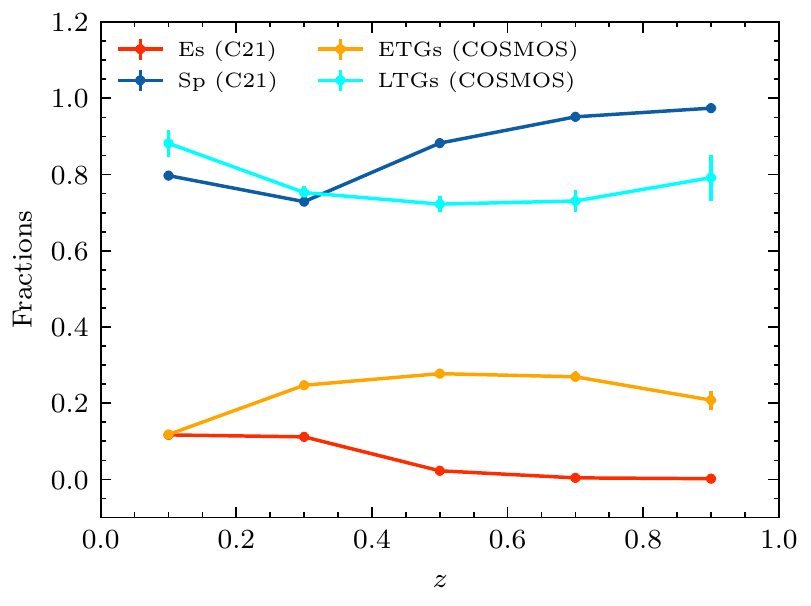}
   	\caption{Fraction of morphological classes as a function of redshift for the V21 (left) and the C21 (right) catalogues. Es/ETG (red) are compared with the galaxies classified as ETG in COSMOS (orange) and Sp/LTG (blue) are compared to galaxies classified as Spirals and Irregulars in COSMOS (cyan). For the V21 sample the results are shown for the full (solid lines) and the secure (dashed lines) samples (defined in Section~\ref{sec:certain_type}). COSMOS galaxies are selected in a similar way as the corresponding morphological catalogue: 6\,909 galaxies have $r$ < 21.5 mag and R$_e$ > 2.8 pixels when compared to V21, and 10\,350 galaxies have $i$ < 21.5 mag and no cut in radius when compared to C21. Vertical error bars are computed assuming Poisson error (i.e., considering that the error in the number of galaxies is $\sqrt{\rm{N}}$).}
    \label{fig:frac_type_vs_z}
\end{center}
\end{figure*}

Binary classification models return a probability value of belonging to a particular class. In order to assign a class, a probability threshold must be defined. Each catalogue has its own recommended thresholds to determine morphology classes, which is referred to as `certain' classification throughout the paper. C21 provides the mean value of the predicted probabilities from five individual models. For each galaxy, an output probability of being an elliptical, $p$(Es), and a spiral, $p$(Sp), is assigned by the C21 machine (with $p$(Es) + $p$(Sp) $=1$). C21 uses a threshold  of 0.8 ($\overline{P}\ge$0.8) to determine if a galaxy is classified as Es or Sp\footnote{C21 also defined a confidence level based on the  S\'ersic index and $g-i$ colour distribution for users who need a further refinement to the classifications that have similar properties in S\'ersic index and colour to the one of the training samples.}. In Fig.~\ref{fig:probability} we show the mean probability distributions of being Sp, $p$(Sp), for the C21 sample. Note that the galaxies in C21 that are not in V21 (labelled as C$\backslash$V) have an asymmetrical probability distribution with a preference for spiral classification. This sub-sample includes a significant fraction of galaxies with a $r$-band radius $\le2.8$ pixels ($\sim40\%$ of C$\backslash$V sample).  


On the other hand, V21 defined `robust' classifications for ETG and LTG when $max\left ( P_{i} \right )<0.3$ and $min\left ( P_{i} \right )>0.7$, respectively, where $P_{i}$ represents the median probability obtained from 5 k-folded models. In addition, galaxies are classified as `secure' when the difference between the maximum and minimum values of the predicted probabilities from the five individual models is smaller than 0.3, i.e. $\triangle P<0.3$. By definition, a `robust' classification is a `secure' classification; galaxies without a robust classification but which satisfies the secure classification criterion are defined as a `secure intermediate' in V21.


For simplicity, throughout the paper, we focus only on the `robust' classifications for V21. For C21, we use classifications based on the probability thresholds mentioned above, i.e. $\overline{P}\ge$0.8. These classifications are referred to as `certain' type in the discussion. The rest of the galaxies not included in any of these selections are categorised as `uncertain' type in this paper\footnote{Note that, by doing so,  the `secure intermediate' galaxies from V21 are not discussed in this paper.}. Hereafter, to distinguish the results from the two catalogues, we will refer to Sp/Es for C21 classifications and to ETG/LTG for V21 ones.


\section{Statistical Comparison}
\label{sec:comp_statistics}

Table~\ref{tab:statistics} reports the number of galaxies of each morphological type in the two catalogues and their different combinations. Due to the different selection criteria used to construct the catalogue samples (Section~\ref{sec:data_selection}), C21 and V21 only partially overlap (C$\cap$V/C$\cup$V$=\sim0.59$). Note that the fraction of Sp/LTG is much larger than the Es/ETG in both catalogues (73\% and 83\% for V21 and C21, respectively). 


Fig.~\ref{fig:frac_type_vs_mag} shows the fraction of each morphological class in magnitude bins (both versus $r$-band and $i$-band magnitudes due to the different sample selection).  The two catalogues are dominated by  disk-like galaxies (i.e., Sp/LTG) at fainter magnitudes. Faint observed galaxies in our sample can be either low mass systems or at high redshift, more likely consistent with the Sp/LTG or peculiar galaxies, which are are likely to be classified as Sp/LTG in binary classifications, based on past studies \citep[e.g.,][etc]{Butcher1984,Dressler1994,Barnes1996,Martel1998,Conselice2005,Conselice2014}.

However, there is a clear difference between the two catalogs at bright magnitudes: in V21, the sample is dominated by ETG up to $r\sim17$, where the trend reverses at fainter magnitudes. On the other hand,  the C21 sample is dominated by Sp at all magnitude bins, and this fraction keeps increasing with magnitude, reaching 94$\%$ at $i\sim20.5$.

To shed more light on this difference, we  compare the results with previous work. In particular, we use the morphological classifications of the COSMOS sample based on support vector machine (SVM) derived by \cite{Huertas-Company2008} and presented in \cite{Tasca2009}. This SVM method classifies the COSMOS galaxies into three classes: Early Type, Spirals and Irregulars. For comparison, we group the galaxies classified as Spirals and Irregulars as LTG. In order to limit the selection effects in the comparison, we report the fractions for COSMOS galaxies with a DES counterpart and use the magnitude and radius measurements from DES. Fig.~\ref{fig:frac_type_vs_z} shows the comparison of the fraction of ETG/LTG according to V21 (left) and Es/Sp according to C21 (right), as a function of redshift, with the ones from the COSMOS sample. The number of COSMOS galaxies that satisfy the V21 criteria (e.g., magnitude and radius limits) is 6\,909. When only the magnitude limit is imposed as in C21, the number of COSMOS galaxies increases to 10\,350. The fraction of LTG for V21 and COSMOS are very consistent in the full redshift range, while the fraction of ETG is somewhat smaller for V21, especially at intermediate redshifts ($z \sim 0.5$). On the other hand, the fraction of Sp from C21 is above that from COSMOS (and the opposite for the Es), especially at higher redshift ($z > 0.3$). We note however that the area covered by COSMOS is very small (2 deg$^2$) compared to the DES one (5\,000 deg$^2$) and thus the results could be severely affected by cosmic variance. In addition, the morphological classification from COSMOS may suffer from their own uncertainties.

In any case, it is evident that there is an overabundance of Sp (reported by C21) compared to LTG (reported by V21). One possibility is that the C21 model, which was trained with bright galaxies only, is confusing noise with features in the faint and blurred images and therefore is over-predicting the number of Sp. More investigation is carried out in Section~\ref{sec:compare_w_DS18}, Section~\ref{sec:bright_mismatch}, and Section~\ref{sec:faint_galaxies}.

\section{Morphologies Agreement \& Disagreement}
\label{sec:examination}
\begin{figure}{}
\begin{center}
\graphicspath{}
	\includegraphics[width=\columnwidth]{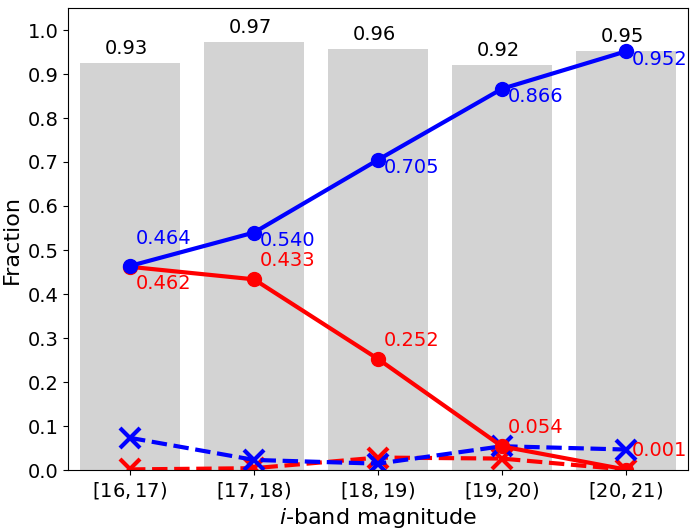}
   	\caption{Agreement of certain types (Section~\ref{sec:certain_type}) within different magnitude bins in $i$-band. Grey bars show the percentage agreement of all galaxies with certain classifications. Solid lines represent the fraction of galaxies with matched classifications between the two catalogues. Blue colour shows the match of Sp \& LTG while red colour is for Es \& ETG. On the contrary, dashed lines and cross points are the fraction of galaxies with mismatched classifications. Blue colour shows the fraction of galaxies that are Sp in C21 but ETG in V21 while red colour present the fraction of ones with a class of Es in C21 and LTG in V21.}
    \label{fig:agreement_hist}
\end{center}
\end{figure}
\begin{figure*}{}
\begin{center}
\graphicspath{}
	\includegraphics[width=2.1\columnwidth]{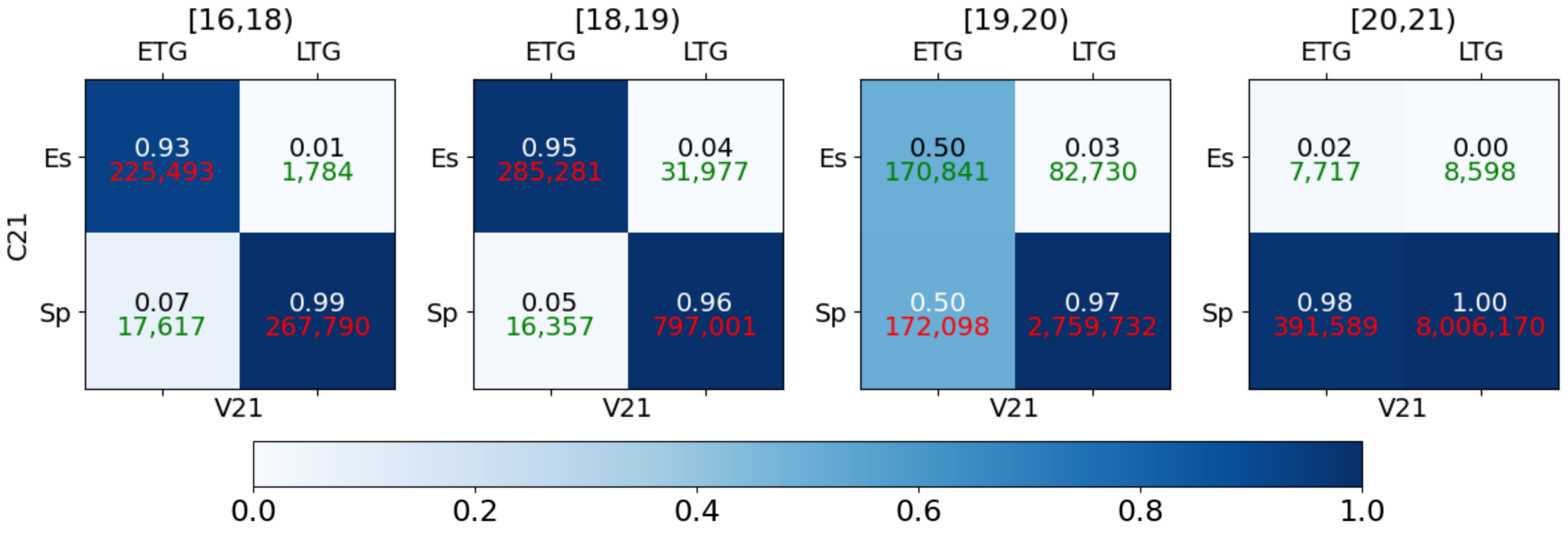}
   	\caption{Confusion matrices of C$\cap$V samples in different magnitude bins. The red or green text in each quadrant represents the number of galaxies with classification in agreement between  C21 and V21. The number above is the fraction of these galaxies  with respect to the V21 classification. }
    \label{fig:label_cm}
\end{center}
\end{figure*}


In this section we study the agreement between the morphological classifications presented in the two catalogues. To be able to compare them one-to-one, we restrict the analysis to the intersection of the two catalogues (17,821,250 galaxies; C$\cap$V in Table~\ref{tab:statistics}). We define the agreement as the fraction of  matched classifications (i.e. Es \& ETG or Sp \& LTG) from the total number of galaxies with a certain classification (Section~\ref{sec:certain_type}). The result is shown in Fig.~\ref{fig:agreement_hist} as a function of $i$ magnitude (used here for convenience since C21 has both lower and upper limits in $i$-band magnitude). The overall agreement is very good, larger than 92$\%$ in all magnitude ranges.


However, while the agreement at bright magnitudes is due both to Es/ETG and Sp/LTG (contributing roughly equally to the overall agreement at 16 $< i <$ 17), at fainter magnitudes, the agreement is largely driven by the high number of Sp/LTG: in the last magnitude bin, $i$=[20,21), only $\sim$0.1$\%$ of the galaxies with consistent morphological type between the two catalogues are classified as Es/ETG.


To further investigate the agreement of each morphology class, we show in Fig.~\ref{fig:label_cm} the confusion matrices in 4 magnitude bins. There is an excellent agreement between both morphology classes up to $i\le19$. Assuming that the V21 classification is correct\footnote{The V21 catalogue performance is discussed in detail when applied to the emulated galaxies, but also with real faint DES galaxies in section 5 of V21.}, this indicates that the CNN predictions of C21  are reliable for samples at least one magnitude fainter than the training sample.

However, at $i > 19$ the agreement vanishes, mostly due to an increasing number of Sp galaxies identified by the C21 catalogue (also see Fig.~\ref{fig:frac_type_vs_mag} and \ref{fig:frac_type_vs_z}). Fig.~\ref{fig:agreement_hist} shows that at $i$=[16,17), about 46$\%$ of the galaxies are classified as ETG by V21 and Es by C21; the percentage significantly drops, to $\sim25\%$, at $i$=[18,19), and the fraction is almost negligible (0.1$\%$) at $i$=[20,21). On the other hand, most of the galaxies classified as LTG in V21 are also classified as Sp by C21. 

The differences in the classification of the faint end galaxies are probably due to the different training strategies: recall that the C21 machine is trained with the images of bright galaxies ($16\le{i}<18$) while V21 included emulated faint DES galaxies up to 22.5 mag in the training sample. In addition, it could also be impacted by the different morphological labels used for training in the two works. Unfortunately, the absence of `ground truth' for the faint DES galaxies prevents us from claiming which are the `right' or `wrong' classifications. In the next sections \ref{sec:bright_mismatch}, \ref{sec:mismatch_faint_galaxies}, and \ref{sec:comp_physics} we investigate in detail the properties of the galaxies with mismatched classifications to shed some light on their nature and their true morphological class.

\subsection{Bright Galaxies}
\label{sec:bright_galaxies}

The previous section suggests that the  comparison of the morphological classifications for the bright and faint galaxies is very different, and so we divide the discussion in the two regimes. In this section, we validate the classifications of bright galaxies ($16\le{i}<18$) taking advantage of the available labels. We compare the classifications with the labels used for training each of the data sets (i.e., GZ1, Section~\ref{sec:compare_w_GZ1}; and DS18, Section~\ref{sec:compare_w_DS18}) and we discuss the nature of the mismatched cases in Section~\ref{sec:bright_mismatch}.

\subsubsection{Compared with GZ1 labels}
\label{sec:compare_w_GZ1}

First, we compare the classifications of the bright galaxies ($16\le{i}<18$) with the labels from Galaxy Zoo 1, used for training C21 models. Both catalogues show an excellent agreement with the GZ1 labels, with less than 3$\%$ of misclassifications.  Interestingly, the dominant mismatches of each catalogue are different. In C21, the main mismatch occurs for galaxies classified as Sp in C21 but labelled as Es in GZ1. C21 and \citet{Cheng2020a} discussed that DES imaging data has better resolution and imaging depth than SDSS revealing  structures such as spiral arms which are not clearly visible in SDSS imaging data, which could explain the disagreement in the classifications. 
Randomly selected examples of this case are shown in Fig.~\ref{fig:example_S+GZ1_Sp+Es}. These galaxies are visually disky and/or spiral galaxies, indicating that C21 classifies these galaxies correctly.


On the other hand, the $\sim$2$\%$ of the misclassifications in V21 correspond to galaxies classified as ETG in V21 but labelled as Sp by GZ1. This mismatch is mostly due to lenticular or edge-on galaxies\footnote{V21 also provides a classification of edge-on galaxies in their catalogue and warns the users not to trust galaxies classified both as ETG and edge-on. Visual inspection confirmed that many of these could be lenticular galaxies seen edge-on.}. Randomly selected examples of this case are shown in Fig.~\ref{fig:example_V+GZ1_ETG+Sp}


\subsubsection{Compared with DS18}
\label{sec:compare_w_DS18}

\begin{figure*}{}
\begin{center}
\graphicspath{}
	\includegraphics[width=2.1\columnwidth]{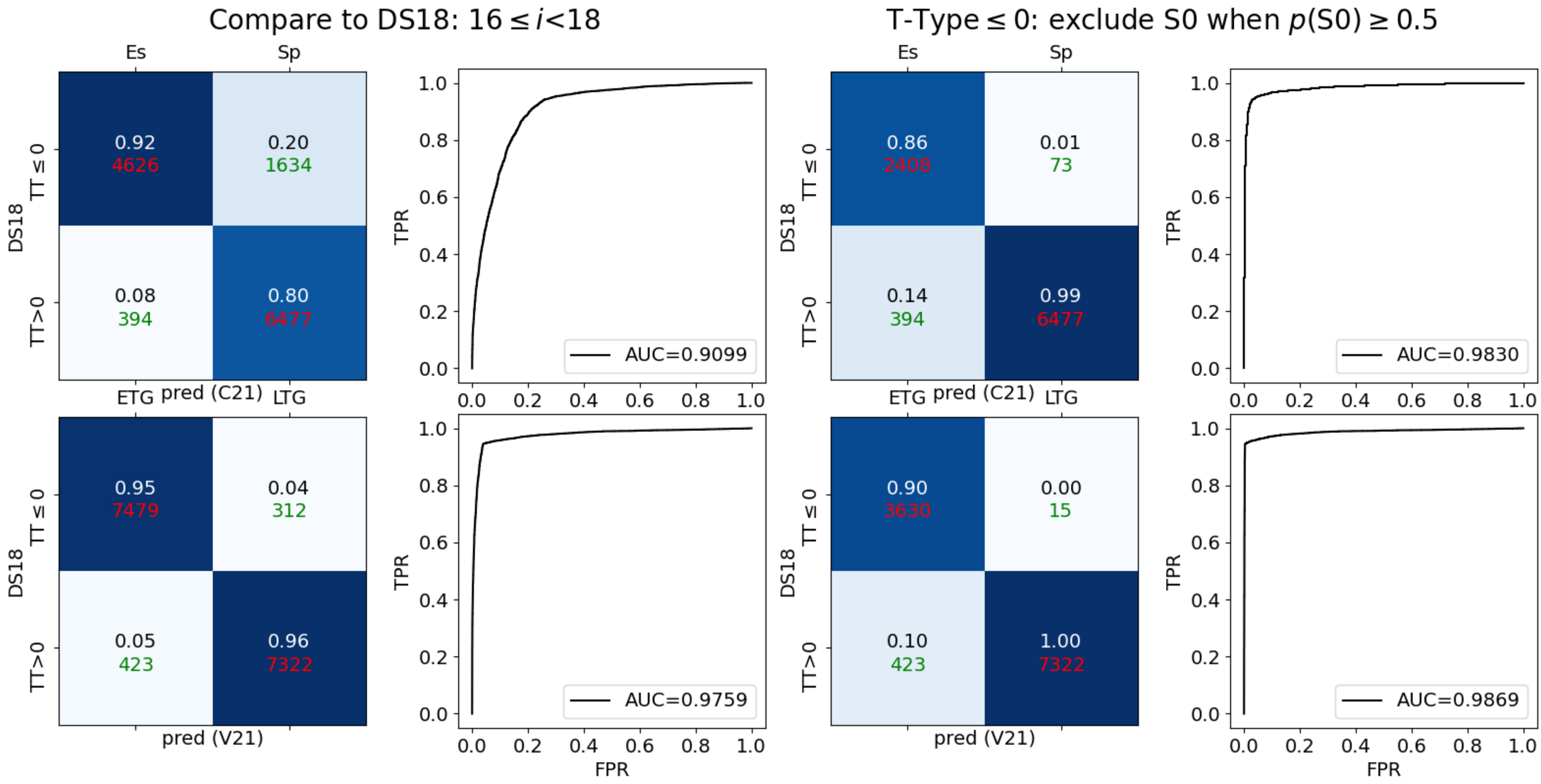}
   	\caption{Confusion matrices and ROC curves of bright galaxies ($16\le{i}<18$) in C21 (top panel) and V21 (bottom panel) compared with DS18. `TT' represents `T-Type'. TT$\le$0 and TT>0 represents ETG and LTG in DS18, respectively. The right panel shows the confusion matrices that exclude lenticular galaxies (S0), i.e. galaxies with T-Type $\le0$ and the probability of being S0, $p$(S0) $\ge0.5$ from DS18.}
    \label{fig:TT_label_check}
\end{center}
\end{figure*}


Next, we compare both catalogues with the labels from DS18 in Fig.~\ref{fig:TT_label_check}. V21 reaches an accuracy of 0.95 while C21 has an accuracy of 0.85, with the main source of mismatches being galaxies classified as Sp in C21 but labeled as ETG (T-Type$\le$0) in DS18.
We check whether this mismatch is dominated by lenticular galaxies using the probability of being lenticular ($p$(S0)) reported in DS18. We confirm that most (0.95) of the galaxies classified as Sp by C21 and labelled as ETG by DS18 are, in fact, lenticular galaxies, consistent with the previous section. Randomly selected examples of this kind of mismatch are shown in Fig.~\ref{fig:example_S+DS18_Sp+ETG}. In addition, we notice that a few of these galaxies show spiral arms. This is the case where the better quality of the DES images reveals structures that are not visible in SDSS, and the corrected classifier in C21 is able to classify them correctly. In the right panel of Fig.~\ref{fig:TT_label_check} we check the role played by lenticular galaxies (classified as such in DS18 according to $p$(S0)) by excluding them from the computation of the confusion matrix. The agreement with C21 classification is significantly improved for the Sp populations, reaching 99$\%$ agreement for this class.

The excellent agreement between the two catalogues and their respective training labels suggests that the inclusion of the three band images does not improve the classification significantly, at least in the bright regime. This can also be important for the construction of future morphological catalogues where single band images can be used.




\subsubsection{Mismatched classifications of bright galaxies}
\label{sec:bright_mismatch}


\begin{figure*}{}
\begin{center}
\graphicspath{}
	\includegraphics[width=2.1\columnwidth]{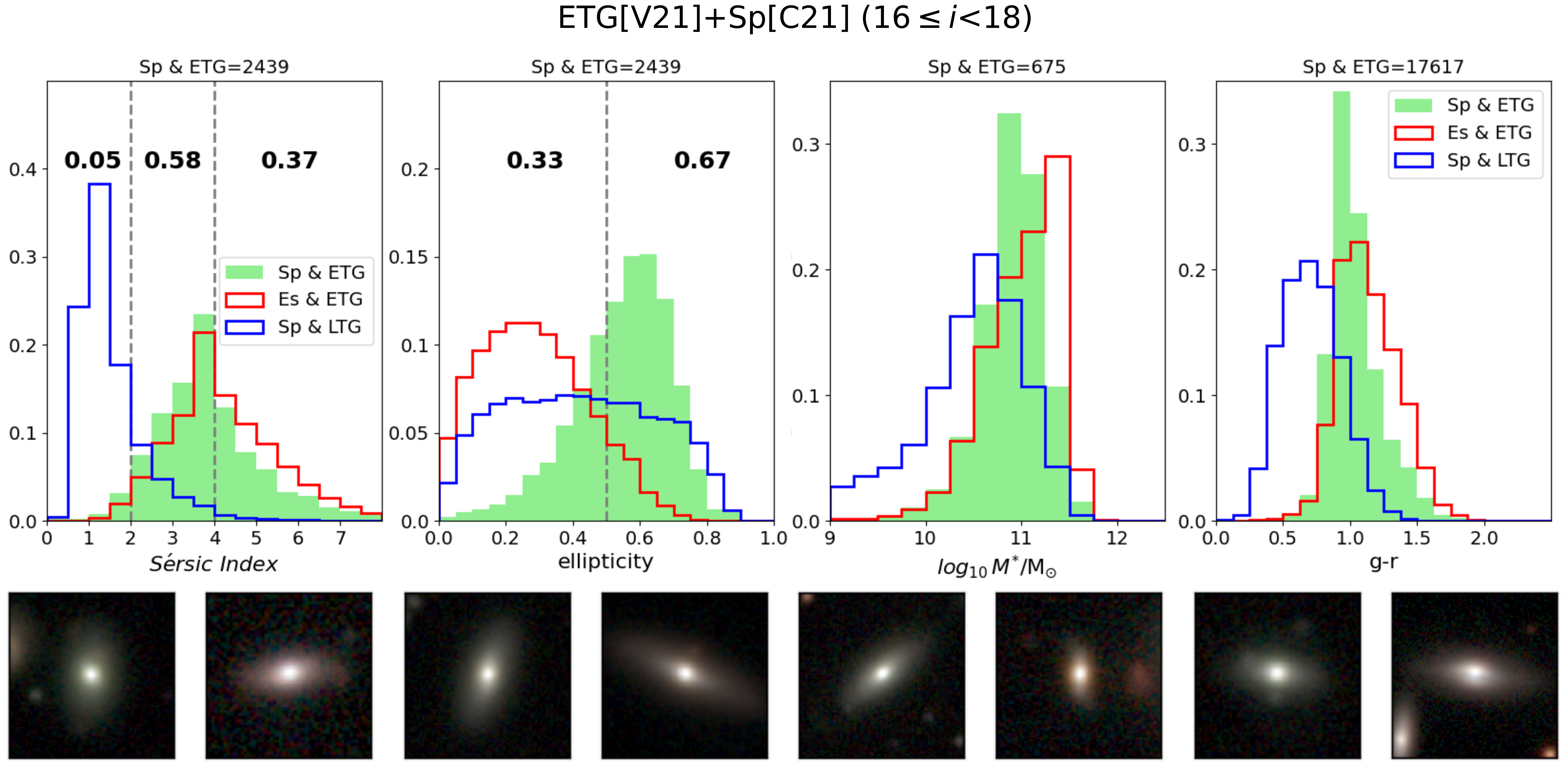}
   	\caption{{\it Top:} the normalised distribution of S\'ersic index, ellipticity, stellar mass, and apparent colour ($g-r$) from left to right. The light green histograms represent Sp \& ETG while the red and blue lines show the case of Es \& ETG and Sp \& LTG, respectively, for comparison. The grey dashed vertical lines note S\'ersic index = 2.0 and 4.0 and ellipticity = 0.5. The text within each area separated by the grey dashed lines shows the fraction of Sp \& ETG in each range. {\it Bottom:} randomly selected example images of the mismatched case, Sp \& ETG.}
    \label{fig:S+V_S0_check}
\end{center}
\end{figure*}

\begin{figure*}{}
\begin{center}
\graphicspath{}
	\includegraphics[width=2\columnwidth]{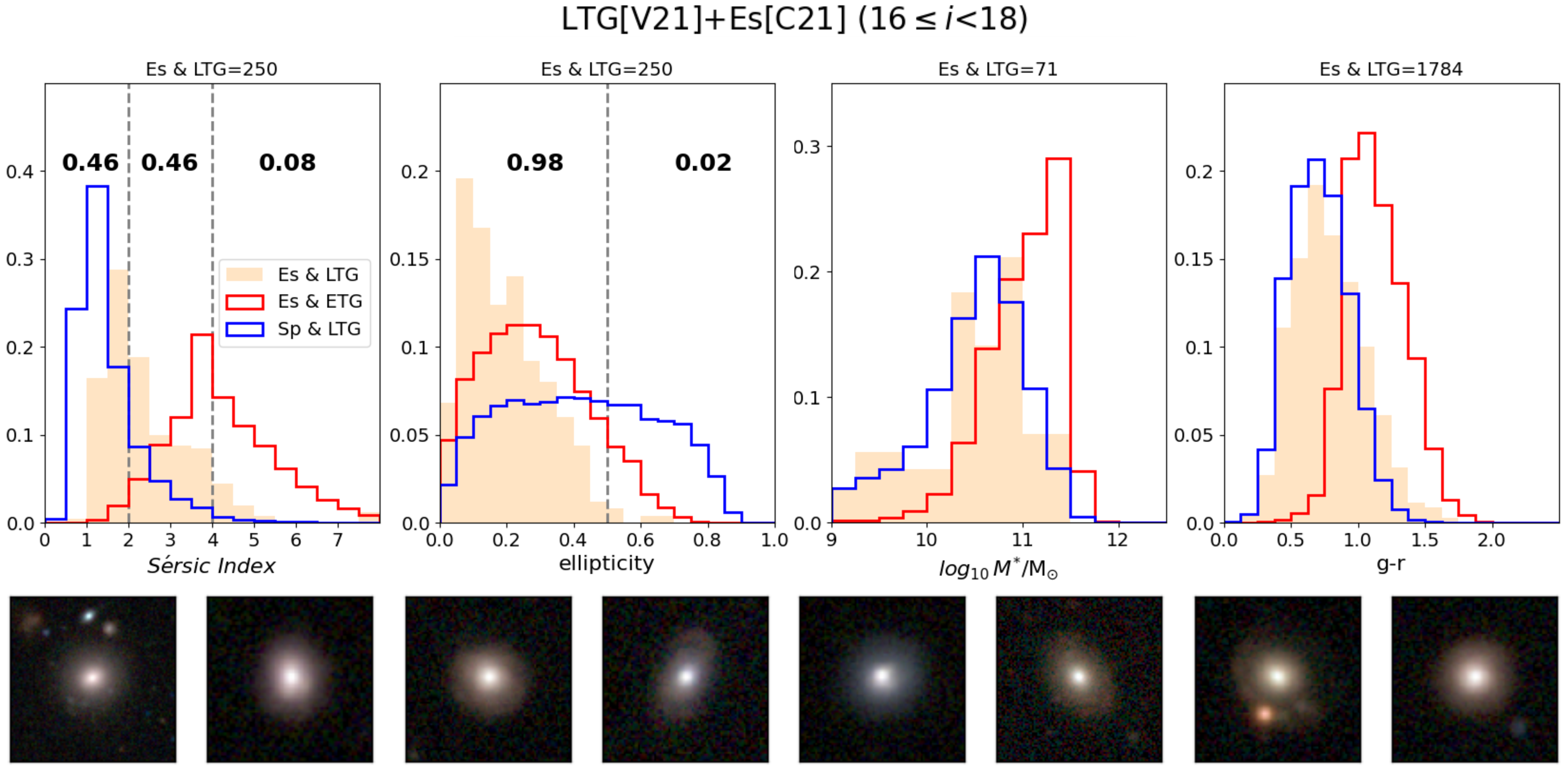}
   	\caption{ Same as Fig. \ref{fig:S+V_S0_check} but for the  Es \& LTG, shown as  light orange histogram.}
    \label{fig:S+V_Es+LTG_bright}
\end{center}
\end{figure*}


We now study in more detail the misclassifications between the two catalogues at $16\le{i}<18$. There are 17,617 Sp \& ETG and 1,784 Es \& LTG, corresponding to $\sim$3.44\%  and $\sim$0.348\% of the certain types within this magnitude range, respectively.

For the case of Sp \& ETG, 0.94 of them are labelled as lenticular galaxies according to DS18. This suggests that bright galaxies with a classification of Sp \& ETG are likely  lenticular galaxies, in agreement with the finding in Section~\ref{sec:compare_w_DS18}. Fig.~\ref{fig:S+V_S0_check} shows  structural and physical properties of
this mismatch. The distributions indicate that
Sp \& ETG galaxies have intermediate to large Sersic index,  high stellar mass, and similar colour distributions to the Es \& ETG. On the contrary, they have a very different ellipticity distribution than Es \& ETG, peaking at elongated values. Randomly selected example images are shown below.




By combining the classifications of the two catalogues and looking for Sp \& ETG, we are able to `re-discover' lenticular galaxies, a class which was not included in any of the training labels used in the construction of the two catalogues. Note, however, that completeness is not ensured when selecting a sample of lenticular galaxies with this approach.


For the  Es \& LTG, we show the S\'ersic index, ellipticity, apparent colour ($g-r$), and stellar mass distribution in Fig.~\ref{fig:S+V_Es+LTG_bright}. Galaxies with a classification of Es \& LTG follow the same colour and mass distribution of Sp \& LTG and tend to have intermediate S\'ersic index, 1 $ < n < $  4, peaking at $n \sim$ 1.5. In addition, they have low ellipticity, i.e., are round objects, suggesting that these galaxies could be disk galaxies seen face-on. Randomly selected examples are shown in Fig.~\ref{fig:S+V_Es+LTG_bright}. Note that these galaxies have a large bulge component, and the spiral structure, if present, is very faint. These galaxies probably correspond to  intermediate types - some of them could be S0 or even Es; these galaxy types are difficult to categorize in any of the two classes.

\subsection{Faint Galaxies}
\label{sec:faint_galaxies}

Despite the good agreement between the two catalogues at bright magnitudes ($i<19$ in Fig.~\ref{fig:label_cm}), this only represents $14\%$ (2,507,490 galaxies) of the intersection of the two catalogues. The lack of a labelled sample at $i>19$ complicates the comparison. Therefore, in this section, we rely on other quantities that correlate with morphology but not uniquely to validate our results. In particular, we use the structural parameters from T18, stellar masses\footnotemark[1] using similar methods to \citet{Palmese2020}, photometric measurements (colour) from the DES Y3 GOLD catalogue, and absolute colour, stellar mass \citep{Moutard2016a,Moutard2016b}, and spectral classifications \citep{Siudek2018b} from VIPERS. Despite a small statistics, we use VIPERS measurements for the absolute colour at fainter magnitudes, because the measurement of absolute magnitudes in DES is less accurate due to the larger uncertainty of the DES photometric redshifts.

\subsubsection{Sp \& ETG mismatch}

\label{sec:mismatch_faint_galaxies}
\begin{figure*}{}
\begin{center}
\graphicspath{}
	\includegraphics[width=2.1\columnwidth]{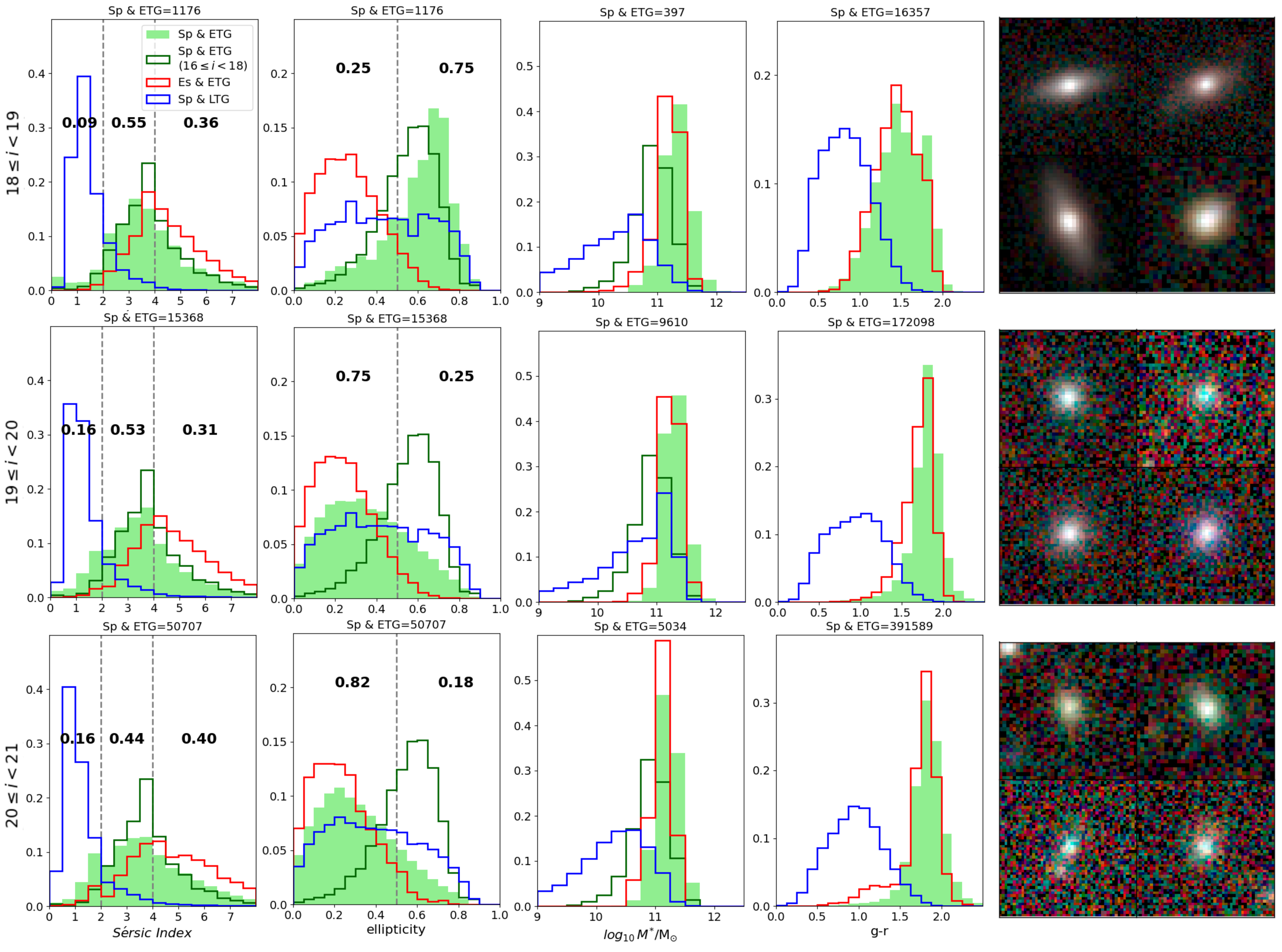}
   	\caption{Normalised distributions of S\'ersic index, ellipticity, stellar mass, and apparent colour, $g-r$, are presented from left to right, and the last panel shows randomly selected examples of Sp \& ETG within each magnitude range. The green shaded histograms show the distribution of Sp \& ETG within each magnitude range. The number of Sp \& ETG galaxies that are used to construct the green shaded histograms is shown above each graph. Solid red and blue lines are the distribution of Es \& ETG and Sp \& LTG, respectively, for comparison. The solid dark green lines show the distributions of the Sp \& ETG case with $16\le{i}<18$ for comparison. The grey vertical dashed lines note S\'ersic index = 2.0 and 4.0 and ellipticity = 0.5. The text within each area separated by the grey lines shows the fraction of  Sp \& ETG in each range. }
    \label{fig:S+V_Sp+ETG_comb}
\end{center}
\end{figure*}
\begin{figure*}{}
\begin{center}
\graphicspath{}
	\includegraphics[width=2.1\columnwidth]{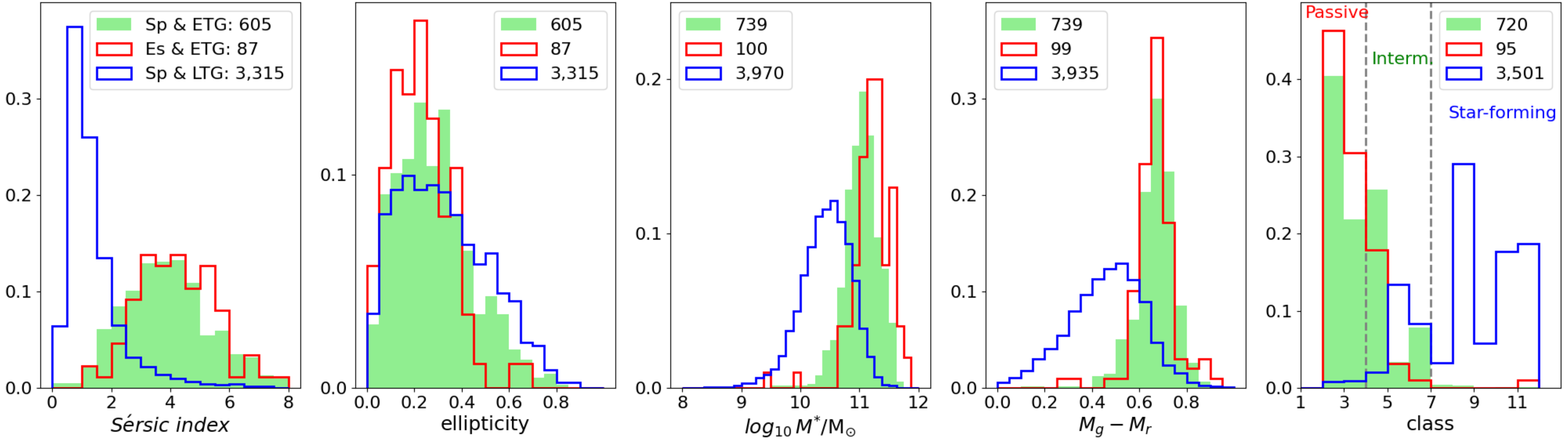}
   	\caption{Normalised distributions of S\'ersic index, ellipticity, stellar mass, absolute colour $M_{g}-M_{r}$ (from VIPERS), and unsupervised spectral classification from \citet{Siudek2018b} are presented. The green shaded histograms show the distributions of Sp \& ETG. The solid red and blue lines are the distributions of Es \& ETG and Sp \& LTG, respectively, for comparison.}
    \label{fig:S+V_Sp+ETG_VIPERS}
\end{center}
\end{figure*}
\begin{figure*}{}
\begin{center}
\graphicspath{}
	\includegraphics[width=2.1\columnwidth]{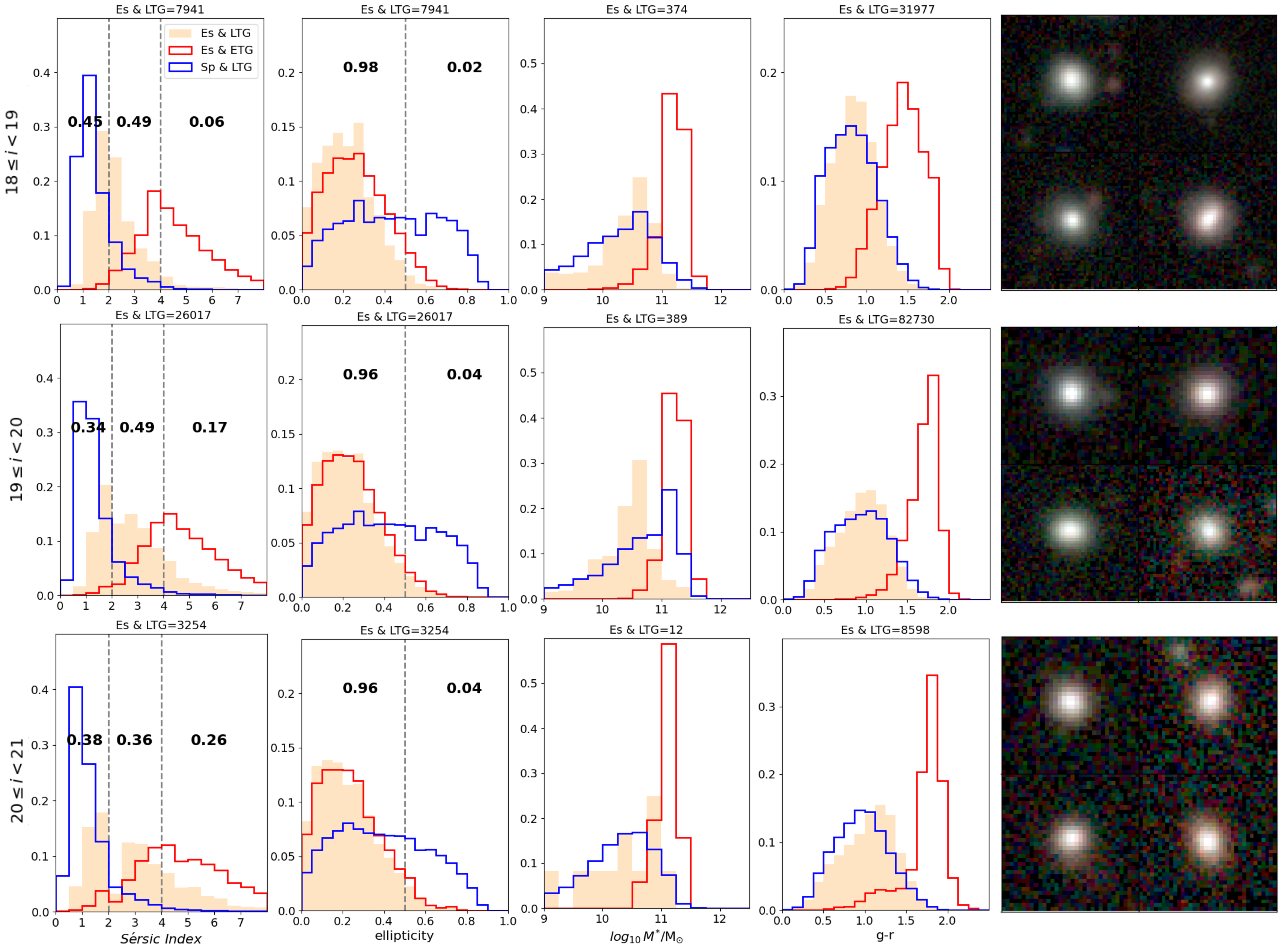}
   	\caption{Same as Fig. \ref{fig:S+V_Sp+ETG_comb} but for the Es \& LTG, shown as orange shaded histograms.}
    \label{fig:S+V_Es+LTG_comb}
\end{center}
\end{figure*}


The number of Sp \& ETG  at $i\ge$18 is 580,044, corresponding only to $\sim$4.6\% of the certain types at $i\ge18$ (as defined in Section~\ref{sec:certain_type}), but as much as $\sim$41\% of the ETG classified as such by V21. Fig.~\ref{fig:label_cm} shows that the agreement between the galaxies classified as ETG drops significantly after $i\ge19$. Hereafter, we separate the discussion into three different magnitude bins: $18\le{i}<19$, $19\le{i}<20$, and $20\le{i}<21$. In Fig.~\ref{fig:S+V_Sp+ETG_comb}, we show the S\'ersic index, ellipticity, stellar mass, and apparent colour ($g-r$) distributions in magnitude bins for the  Sp \& ETG. Note that the colour is based on observed magnitudes, not absolute or k-corrected due to the uncertainty in DES photometric redshift, which complicates the comparison at different redshifts. We use apparent colour distributions of the matched classifications in each magnitude bin as a comparison. 



Galaxies classified as Sp \& ETG follow a similar S\'ersic index distribution of bright Sp \& ETG in all magnitude bins, peaking at intermediate range ($n \sim 3$). They are red and massive objects, with very similar colour and mass distributions to the Es \& ETG. 
Interestingly, the ellipticity ($\epsilon$) distributions are very different at different magnitude ranges: the Sp \& ETG $18\le{i}<19$ have large ellipticities (i.e., they are elongated galaxies), similar to the bright Sp \& ETG, supporting the assumption that they are potential lenticular galaxies. On the other hand, at  ${i}>19$ they span a much wider range peaking at $\epsilon \sim 0.2$ (i.e., rounder objects).




The final panel shows randomly selected examples of Sp \& ETG. While the ones at $18\le{i}<19$ look like lenticulars seen edge-on, the images at fainter magnitudes are very noisy and it is difficult to confirm their morphology by eye. Because of the significant change in properties, we cannot argue that the Sp \& ETG at fainter magnitude are dominated by lenticular galaxies.

In order to further investigate the nature of these Sp \& ETG galaxies, we use measurements from VIPERS, which provide absolute colors and a classification based on spectroscopic information. Unfortunately, the number of galaxies in the two catalogues with a VIPERS counterpart is very small ($\sim4800$ in total, $\sim600-700$ Sp \& ETG, 99\% of them with $i\ge19$), and thus we cannot separate the sample into magnitude bins. Fig.~\ref{fig:S+V_Sp+ETG_VIPERS} shows similar trends to Fig.~\ref{fig:S+V_Sp+ETG_comb}. The distribution of S\'ersic index and absolute colour ($M_g-M_r$) of Sp \& ETG is even more similar to the one of Es \& ETG. In addition, 0.99 of the Sp \& ETG have spectral classifications from \citet{Siudek2018b}\footnote{\citet{Siudek2018b} used a Fisher Expectation-Maximization (FEM) unsupervised algorithm to categorize galaxies in 12 spectral classes.}, consistent with passive or intermediate populations (class=[0,7]). These results suggest that this population comprises distant passive/intermediate massive galaxies with round shapes, more consistent with ETGs, including potential lenticular galaxies (although this is very difficult to say from visual inspection alone due to the noisy images). Note that there are only 4 galaxies  classified as Sp \& ETG at $18\le{i}<19$ with VIPERS's measurements. For $i\ge19$, we also have small statistics $-$ 605 galaxies; thus, for this specific magnitude range more precise measurements to make a robust comparison are needed.

As a side note, the Sp \& LTG population at $19\le{i}<20$ contains more massive galaxies (with a peak at $log_{10}\,M^{*}/{\rm M}_{\odot} \sim 11$) than any other magnitude bin. Since there is not an apparent change in S\'ersic index, ellipticity, and colour, these massive galaxies may be interesting targets to followup.



 \subsubsection{Es \& LTG mismatch}

Now we discuss the Es \& LTG case, which corresponds to 123,305 galaxies ($\sim$1\% of the certain types and $\sim$1\% of the LTG according to V21), i.e, a much smaller fraction than the mismatch Sp \& ETG.

Fig.~\ref{fig:S+V_Es+LTG_comb} shows the S\'ersic index, ellipticity, stellar mass, and apparent colour distributions for  Es \& LTG. In general, these galaxies have similar distributions in mass and colour as the Sp \& LTG population. However, they are round objects with intermediate values of S\'ersic indices. The cutout examples have no signs of spiral or asymmetric structure, suggesting that these could be face-on spiral galaxies with low T-Types (0 $<$ T-Type $<$ 2), or even low mass elliptical galaxies. Again, these are intermediate galaxies which are difficult to classify, similarly to the brighter population discussed in Section~\ref{sec:bright_mismatch}. Note that the fraction of Es \& LTG with high S\'ersic index (i.e. S\'ersic index $>$ 2) increases after $i\ge19$. In particular, a bimodal distribution appears in the fainter bin ($20\le{i}<21$). These galaxies with bulge structure (S\'ersic index $>4$) are blue galaxies with relatively lower mass (peaking at $log_{10}\,M^{*}/{\rm M}_{\odot} \sim 10.5$) compared to the ones of Es \& ETG. Although the S\'ersic index measurements are more uncertain for this kind of galaxies, which complicates the analysis, it is reasonable that these galaxies have uncertain classification since they are uncommon in the initial training samples (i.e. bright galaxies) but also due to their intrinsic faintness, which complicates the morphological classification. This shows that, although emulating bright galaxies to fainter magnitudes for training improves the CNN accuracy, some populations of galaxies could be missing. As an alternative, using hydrodynamic simulations to build complete populations of galaxies throughout the magnitude and redshift ranges of the targets in the training process might help solving this bias. By adopting the certain classifications provided in the two catalogues, one could alternatively train a series of machines that contains real faint galaxies.



\section{Comparison of Galaxies Properties}
\label{sec:comp_physics}
\begin{figure*}{}
\begin{center}
\graphicspath{}
	\includegraphics[width=2.1\columnwidth]{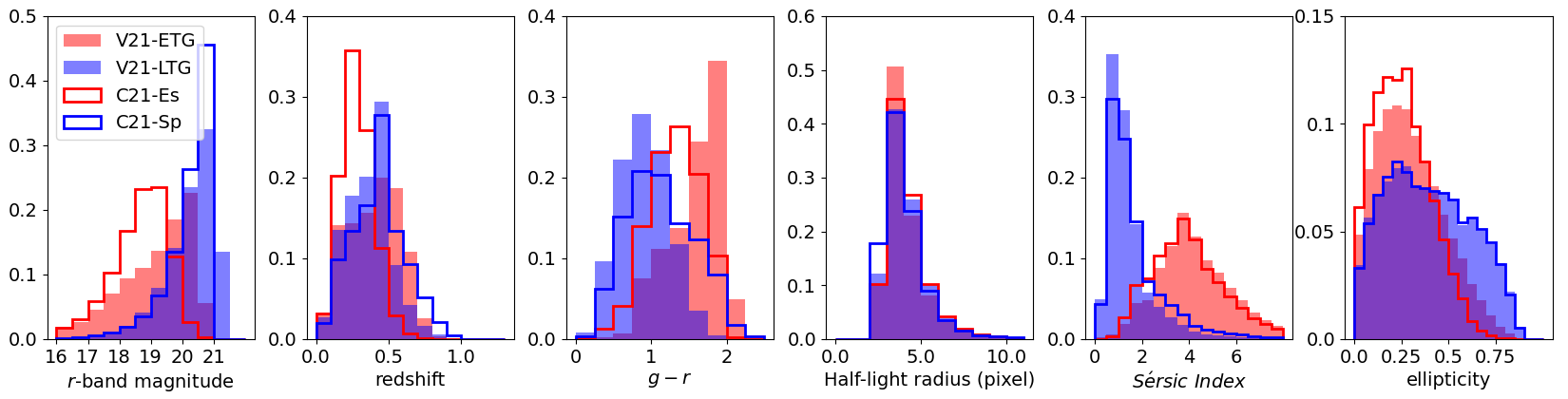}
   	\caption{Normalised distributions of $r$-band magnitude, redshift, apparent colour ($g-r$), half-light radius, S\'ersic index, and ellipticity. The shaded histograms show galaxies with robust classification from V21 while solid lines are for C21. Red and blue colour indicates Es/ETG and Sp/LTG, respectively.}
    \label{fig:alltype_phycheck}
\end{center}
\end{figure*}
\begin{figure}{}
\begin{center}
\graphicspath{}
	\includegraphics[width=\columnwidth]{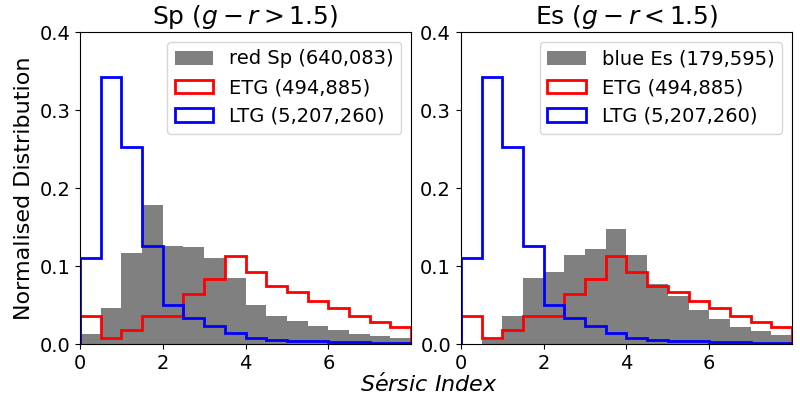}
   	\caption{The shaded histograms show the normalised S\'ersic index distributions of red Sp with $g-r>1.5$ (left) and blue Es with $g-r<1.5$ (right) from C21. The blue and red solid lines show the S\'ersic index distributions of LTG and ETG from V21, respectively, for comparison. The number of data in each class that have S\'ersic index measurements is shown within brackets in the legend.}
    \label{fig:sersic_type_vs_diff_colour}
\end{center}
\end{figure}

Finally, without limiting to the intersection samples, we study several physical and structural properties in each catalogue  for the samples with a certain classification, i.e. either Es or Sp in C21 and either ETG or LTG in V21, in order to further discuss what might impact each machine's decision. Note that the classification criteria used for V21 classifications is the more restrictive version, i.e., robust ETGs/LTGs (Section~\ref{sec:certain_type})\footnote{One can choose to use a less restrictive criteria to increase the statistics, depending on the scientific goals.}.


Fig.~\ref{fig:alltype_phycheck} shows the distributions of magnitude, redshift, observed colour and structural parameters (half-light radius, S\'ersic index, ellipticity) for the two classes (ETG/LTG or Es/Sp) in the two catalogues. While the distributions in size and ellipticity look quite similar for the two catalogues, there are significant differences in the other properties.


In particular, ETG and LTG, classified according to V21, have similar redshift distributions, but show a clear bimodality in colour. The ETG have brighter observed magnitudes compared to LTG but span through the full  magnitude range, at least up to 21 mag in the $r$-band. On the other hand, the colors of Es and Sp galaxies, classified as such according to C21, are more overlapped, while the Es population clearly dominates at bright magnitudes (peaking at $\sim$18.5 mag in the $r$-band) and lower redshifts compared to the Sp population.



We further investigate the discrepancy in colour distribution by studying the S\'ersic index distribution of red Sp ($g-r>1.5$) and blue Es ($g-r<1.5$) in Fig.~\ref{fig:sersic_type_vs_diff_colour}. Note that we use apparent colour in comparison without addressing K-correction; hence, we apply a colour cut at $g-r=1.5$, at the intersection point of the colour distributions of ETG and LTG (Fig.~\ref{fig:alltype_phycheck}). Blue galaxies ($g-r<1.5$) classified as Es in C21 do closely follow the S\'ersic index distribution of ETG from V21. On the other hand, the red ($g-r$>1.5) galaxies classified as Sp have a intermediate S\'ersic index distribution in between ETG and LTG. This indicates a significant structural difference between red Sp and ETG/LTG which may indicate that they are potential lenticular galaxies.

Red Sp are $\sim$21\% of galaxies at $z<1$ with a certain classification in C21 ($\sim$18\% from the total sample). 
In Section~\ref{sec:bright_mismatch}, we discussed that the Sp \& ETG can be potential lenticular galaxies. Over 90\% of the Sp \& ETG are red Sp (but only 22.5\% red Sp are Sp \& ETG). However, we  note again that there are a variety of uncertainties associated with identifying lenticular galaxies, since they cover a wide range of physical properties \citep{Deeley2020,Deeley2021}. We remark that the pursuit in this paper is not to identify lenticular galaxies, but to assess the impact of different CNN approaches to this undefined class.

\section{Summary and conclusions}
\label{sec:summary}

In this work we compare the two largest galaxy morphological classification catalogues to date: V21, which includes $\sim$27 million galaxies at $r<21.5$, and C21, which includes $\sim$21 million galaxies with $16\le{i}<21$ and $z<1$. Due to different initial sample selection, the two catalogues have an overlap of $\sim60\%$. Besides differences in the CNN architectures, the two studies have fundamental differences in their approaches such as (1)  cutout sizes; (2) training labels; (3) brightness of the training sample; and (4) input to the CNN: multi-band vs monochromatic images (Section~\ref{sec:method}). The fact that such different methodologies are applied to the same large datasets, allows us to compare the results in a statistically  significant way and to assess different machine learning approaches.

We examine the agreement between the two catalogues using the intersection sample (C$\cap$V). The agreement is as high as $\sim$95$\%$ (see Fig.~\ref{fig:agreement_hist}) of the galaxies with a certain classification, i.e. Es \& ETG or Sp \& LTG (as defined in Section~\ref{sec:certain_type}). However, the large agreement is mostly driven by the Sp \& LTG population, which corresponds to 66\% of the intersection of the two catalogues (89\% of the samples with a certain classification from one of the two catalogues). On the other hand, the agreement of the Es/ETG population is lower (3.9\%), especially at $i > 19$ mag ($\sim$1.2\% of the total intersection sample), where the number of galaxies classified as Es by C21 decreases (see Fig.~\ref{fig:frac_type_vs_mag}).

The  classifier of C21 was trained with bright low redshift galaxies only, where reliable visual labels were available, while V21 added faint galaxies to the training sample by emulating bright galaxies to fainter magnitudes. One of the main results from this comparison is the fact that the C21 machine can push its predictions accurately one magnitude fainter than its training samples. The excellent agreement between the two catalogues up to ${i}<19$ also indicates that the use of multi-band images does not provide a significant improvement in the morphological classifications of galaxies when compared to the use of monochromatic images. This can have a significant impact on the machine learning methodologies used for morphological classifications in future Big Data surveys such as Euclid or Rubin/LSST.


We have studied in detail the photometric, structural and spectroscopic properties of the mismatched classifications of the two catalogues divided in the bright ($i<18$) and faint ($i\ge18$) regimes. Our main findings are:
\begin{itemize}
\item Bright Sp \& ETG: These are $\sim$3.4$\%$ of the galaxies with certain classification in this magnitude range. Their properties and appearance are consistent with possibly being lenticular galaxies (Fig.~\ref{fig:S+V_S0_check}), a class that was not included in any of the training samples. This mismatch could be a potential way to identify lenticular galaxies, but we note that this approach might result in a small completeness.
\item Bright Es \& LTG: These contribute less than 0.4$\%$ to the  galaxies with certain classification in this magnitude range. They are generally  blue,  round galaxies (95\%  have ellipticity smaller than 0.5, see Fig.~\ref{fig:S+V_Es+LTG_bright}) with intermediate Sèrsic indices. These galaxies are difficult to classify and could be a mixture of face-on disk galaxies with no signs of spiral structure, lenticular, or elliptical galaxies.
\item Faint Sp \& ETG: This population contributes  to $\sim$5\% of the galaxies with certain classification in this magnitude range but as much as 41\% of the ETG classified as such by V21. At $18\le{i}<19$, their structural measurements and physical properties follow similar trends as the ones at $16\le{i}<18$, supporting the hypothesis that  they could be lenticular galaxies as well. In the fainter magnitude bins (${i}>19$) their properties are more consistent with a red, massive, and passive population, although it is almost impossible to confirm via visual inspection due to their noisy images (see Fig.~\ref{fig:S+V_Sp+ETG_comb}).

\item Faint Es \& LTG: These correspond to $\sim$ 1\% of the galaxies with a certain classification for both catalogs in this magnitude range. These systems are blue galaxies with relatively low masses compared to the Es \& ETG, with round shapes, and intermediate Sèrsic indices. These galaxies are difficult to classify, as in the bright regime (Fig.~\ref{fig:S+V_Es+LTG_comb}). At $i\ge19$, there is a bimodal distribution of the S\'ersic index, i.e., they could be a population of blue low mass galaxies with significant bulge component, which may not be well-represented in the training sample. 
\end{itemize}

Finally, we compare the physical properties of the galaxies classified as Es/ETG and Sp/LTG from each catalogue in Fig.~\ref{fig:alltype_phycheck}.  The galaxies classified as ETG by V21 have similar structural measurements such as half-light radius, S\'ersic index, and ellipticity to the galaxies classified as Es by C21; the same is true for Sp/LTG. However, the distribution in observed magnitude, redshift and colour are very different for the C21 compared to V21. ETG and LTG galaxies span more or less over the full magnitude and redshift range but are clearly separated in colour, with ETGs being redder. On the other hand, the colors of Es/Sp are more overlapped and instead there is a clear difference in the magnitude and redshift of Es and Sp, with Es being more abundant in the bright low redshift regime. This suggests that the C21 classifier underpredicts Es at fainter magnitudes and high redshifts (see also Fig.~\ref{fig:frac_type_vs_z}) probably because it confuses noise with structure. However, interestingly, when examining blue Es ($g-i<1.5$) and red Sp ($g-i>1.5$) in Fig.~\ref{fig:sersic_type_vs_diff_colour}, blue Es have similar structure to ETG and red Sp show intermediate structure between ETG and LTG.


This is the first comparison of two largest morphological catalogues up to date, including over 20 million galaxies each, generated by a CNN. This allows us to assess different CNN approaches and to further validate the classifications provided by each catalogue. These classifications could indeed serve as a training sample to classify future datasets  at faint magnitude limits. In the case of disagreement between the classifications ($\sim5$\% of intersection of the two samples), we suggest to complement the classification with additional parameters, such as structural measurements. Furthermore, by combining the two catalogues, the number of total classifications is increased up to $\sim$30 million galaxies for future studies of galaxy evolution. We warn the reader to be aware of the biases and disagreement detailed in this paper when making use of the catalogues for scientific research. In particular, we caution on the robustness of the selection of the ETG population at faint magnitudes due to the large discrepancy of the classifications presented in the two catalogues.



\section*{Acknowledgements}
We thank Dr Thibaud Moutard for providing the measurements of galaxy stellar mass in VIPERS to this work. TYC acknowledges the support of STFC grant ST/T000244/1 and Royal Society grant RF/ERE/210326, hosted at Durham University. HDS acknowledges support by the PID2020-115098RJ-I00 grant from MCIN/AEI/10.13039/501100011033. MS has been supported by the European Union's Horizon 2020 Research and Innovation programme under the Maria Sklodowska-Curie grant agreement (No. 754510) and the Spanish Ministry of Science and Innovation through the Juan de la Cierva-formacion programme (FJC2018-038792-I). MS has been supported by the Polish National Agency for Academic Exchange (Bekker grant BPN/BEK/2021/1/00298/DEC/1).


Funding for the DES Projects has been provided by the U.S. Department of Energy, the U.S. National Science Foundation, the Ministry of Science and Education of Spain, the Science and Technology Facilities Council of the United Kingdom, the Higher Education Funding Council for England, the National Center for Supercomputing Applications at the University of Illinois at Urbana-Champaign, the Kavli Institute of Cosmological Physics at the University of Chicago, the Center for Cosmology and Astro-Particle Physics at the Ohio State University, the Mitchell Institute for Fundamental Physics and Astronomy at Texas A$\&$M University, Financiadora de Estudos e Projetos, Fundação Carlos Chagas Filho de Amparo à Pesquisa do Estado do Rio de Janeiro, Conselho Nacional de Desenvolvimento Científico e Tecnológico and the Ministério da Ciência, Tecnologia e Inovação, the Deutsche Forschungsgemeinschaft, and the Collaborating Institutions in the Dark Energy Survey.

The Collaborating Institutions are Argonne National Laboratory, the University of California at Santa Cruz, the University of Cambridge, Centro de Investigaciones Energéticas, Medioambientales y Tecnológicas-Madrid, the University of Chicago, University College London, the DES-Brazil Consortium, the University of Edinburgh, the Eidgenössische Technische Hochschule (ETH) Zürich, Fermi National Accelerator Laboratory, the University of Illinois at Urbana-Champaign, the Institut de Ciències de l'Espai (IEEC/CSIC), the Institut de Física d'Altes Energies, Lawrence Berkeley National Laboratory, the Ludwig-Maximilians Universität München and the associated Excellence Cluster Universe, the University of Michigan, the National Optical Astronomy Observatory, the University of Nottingham, The Ohio State University, the University of Pennsylvania, the University of Portsmouth, SLAC National Accelerator Laboratory, Stanford University, the University of Sussex, Texas A$\&$M University, and the OzDES Membership Consortium.

Based in part on observations at Cerro Tololo Inter-American Observatory, National Optical Astronomy Observatory, which is operated by the Association of Universities for Research in Astronomy (AURA) under a cooperative agreement with the National Science Foundation.

The DES data management system is supported by the National Science Foundation under grant numbers AST-1138766 and AST-1536171. The DES participants from Spanish institutions are partially supported by MINECO under grants AYA2015-71825, ESP2015-66861, FPA2015-68048, SEV-2016-0588, SEV-2016-0597, and MDM-2015-0509, some of which include ERDF funds from the European Union. IFAE is partially funded by the CERCA programme of the Generalitat de Catalunya. Research leading to these results has received funding from the European Research Council under the European Union's Seventh Framework Program (FP7/2007-2013) including ERC grant agreements 240672, 291329, and 306478. We acknowledge support from the Australian Research Council Centre of Excellence for All-sky Astrophysics (CAASTRO), through project number CE110001020, and the Brazilian Instituto Nacional de Ciencia e Tecnologia (INCT) e-Universe (CNPq grant 465376/2014-2).


\section*{Data Availability}
The morphological catalogue of V21 and C21 are available in the Dark Energy Survey Data Management (DESDM) system at the National Center for Supercomputing Applications (NCSA) at the University of Illinois and can be accessed at \url{https://des.ncsa.illinois.edu/releases/other/morphCNN}.

\bibliographystyle{mnras}
\bibliography{ms}

\begin{thebibliography}{}
\makeatletter
\relax
\def\mn@urlcharsother{\let\do\@makeother \do\$\do\&\do\#\do\^\do\_\do\%\do\~}
\def\mn@doi{\begingroup\mn@urlcharsother \@ifnextchar [ {\mn@doi@}
  {\mn@doi@[]}}
\def\mn@doi@[#1]#2{\def\@tempa{#1}\ifx\@tempa\@empty \href
  {http://dx.doi.org/#2} {doi:#2}\else \href {http://dx.doi.org/#2} {#1}\fi
  \endgroup}
\def\mn@eprint#1#2{\mn@eprint@#1:#2::\@nil}
\def\mn@eprint@arXiv#1{\href {http://arxiv.org/abs/#1} {{\tt arXiv:#1}}}
\def\mn@eprint@dblp#1{\href {http://dblp.uni-trier.de/rec/bibtex/#1.xml}
  {dblp:#1}}
\def\mn@eprint@#1:#2:#3:#4\@nil{\def\@tempa {#1}\def\@tempb {#2}\def\@tempc
  {#3}\ifx \@tempc \@empty \let \@tempc \@tempb \let \@tempb \@tempa \fi \ifx
  \@tempb \@empty \def\@tempb {arXiv}\fi \@ifundefined
  {mn@eprint@\@tempb}{\@tempb:\@tempc}{\expandafter \expandafter \csname
  mn@eprint@\@tempb\endcsname \expandafter{\@tempc}}}

\bibitem[\protect\citeauthoryear{{Abbott} et~al.,}{{Abbott}
  et~al.}{2018}]{Abbott2018}
{Abbott} T.~M.~C.,  et~al., 2018, \mn@doi [\apjs] {10.3847/1538-4365/aae9f0},
  \href {https://ui.adsabs.harvard.edu/abs/2018ApJS..239...18A} {239, 18}

\bibitem[\protect\citeauthoryear{{Alarcon} et~al.,}{{Alarcon}
  et~al.}{2021}]{Alarcon2021}
{Alarcon} A.,  et~al., 2021, \mn@doi [\mnras] {10.1093/mnras/staa3659}, \href
  {https://ui.adsabs.harvard.edu/abs/2021MNRAS.501.6103A} {501, 6103}

\bibitem[\protect\citeauthoryear{{Arnouts} \& {Ilbert}}{{Arnouts} \&
  {Ilbert}}{2011}]{Arnouts2011}
{Arnouts} S.,  {Ilbert} O.,  2011, {LePHARE: Photometric Analysis for Redshift
  Estimate} (\mn@eprint {ascl} {1108.009})

\bibitem[\protect\citeauthoryear{{Baillard} et~al.,}{{Baillard}
  et~al.}{2011}]{Baillard2011}
{Baillard} A.,  et~al., 2011, \mn@doi [\aap] {10.1051/0004-6361/201016423},
  \href {https://ui.adsabs.harvard.edu/abs/2011A&A...532A..74B} {532, A74}

\bibitem[\protect\citeauthoryear{{Banerji} et~al.,}{{Banerji}
  et~al.}{2010}]{Banerji2010}
{Banerji} M.,  et~al., 2010, \mn@doi [\mnras]
  {10.1111/j.1365-2966.2010.16713.x}, \href
  {https://ui.adsabs.harvard.edu/abs/2010MNRAS.406..342B} {406, 342}

\bibitem[\protect\citeauthoryear{{Barnes} \& {Hernquist}}{{Barnes} \&
  {Hernquist}}{1996}]{Barnes1996}
{Barnes} J.~E.,  {Hernquist} L.,  1996, \mn@doi [\apj] {10.1086/177957}, \href
  {https://ui.adsabs.harvard.edu/abs/1996ApJ...471..115B} {471, 115}

\bibitem[\protect\citeauthoryear{{Bottrell} et~al.,}{{Bottrell}
  et~al.}{2019}]{Bottrell2019}
{Bottrell} C.,  et~al., 2019, \mn@doi [\mnras] {10.1093/mnras/stz2934}, \href
  {https://ui.adsabs.harvard.edu/abs/2019MNRAS.490.5390B} {490, 5390}

\bibitem[\protect\citeauthoryear{{Bruzual} \& {Charlot}}{{Bruzual} \&
  {Charlot}}{2003}]{Bruzual2003}
{Bruzual} G.,  {Charlot} S.,  2003, \mn@doi [\mnras]
  {10.1046/j.1365-8711.2003.06897.x}, \href
  {https://ui.adsabs.harvard.edu/abs/2003MNRAS.344.1000B} {344, 1000}

\bibitem[\protect\citeauthoryear{{Butcher} \& {Oemler}}{{Butcher} \&
  {Oemler}}{1984}]{Butcher1984}
{Butcher} H.,  {Oemler} A. J.,  1984, \mn@doi [\apj] {10.1086/162519}, \href
  {https://ui.adsabs.harvard.edu/abs/1984ApJ...285..426B} {285, 426}

\bibitem[\protect\citeauthoryear{{Cheng} et~al.,}{{Cheng}
  et~al.}{2020a}]{Cheng2020a}
{Cheng} T.-Y.,  et~al., 2020a, \mn@doi [\mnras] {10.1093/mnras/staa501}, \href
  {https://ui.adsabs.harvard.edu/abs/2020MNRAS.493.4209C} {493, 4209}

\bibitem[\protect\citeauthoryear{{Cheng}, {Li}, {Conselice},
  {Arag{\'o}n-Salamanca}, {Dye}  \& {Metcalf}}{{Cheng}
  et~al.}{2020b}]{Cheng2020b}
{Cheng} T.-Y.,  {Li} N.,  {Conselice} C.~J.,  {Arag{\'o}n-Salamanca} A.,  {Dye}
  S.,   {Metcalf} R.~B.,  2020b, \mn@doi [\mnras] {10.1093/mnras/staa1015},
  \href {https://ui.adsabs.harvard.edu/abs/2020MNRAS.494.3750C} {494, 3750}

\bibitem[\protect\citeauthoryear{{Cheng}, {Huertas-Company}, {Conselice},
  {Arag{\'o}n-Salamanca}, {Robertson}  \& {Ramachandra}}{{Cheng}
  et~al.}{2021a}]{Cheng2021-uml}
{Cheng} T.-Y.,  {Huertas-Company} M.,  {Conselice} C.~J.,
  {Arag{\'o}n-Salamanca} A.,  {Robertson} B.~E.,   {Ramachandra} N.,  2021a,
  \mn@doi [\mnras] {10.1093/mnras/stab734}, \href
  {https://ui.adsabs.harvard.edu/abs/2021MNRAS.503.4446C} {503, 4446}

\bibitem[\protect\citeauthoryear{{Cheng} et~al.,}{{Cheng}
  et~al.}{2021b}]{Cheng2021-cata}
{Cheng} T.-Y.,  et~al., 2021b, \mn@doi [\mnras] {10.1093/mnras/stab2142}, \href
  {https://ui.adsabs.harvard.edu/abs/2021MNRAS.507.4425C} {507, 4425}

\bibitem[\protect\citeauthoryear{{Collister} \& {Lahav}}{{Collister} \&
  {Lahav}}{2004}]{Collister2004}
{Collister} A.~A.,  {Lahav} O.,  2004, \mn@doi [\pasp] {10.1086/383254}, \href
  {https://ui.adsabs.harvard.edu/abs/2004PASP..116..345C} {116, 345}

\bibitem[\protect\citeauthoryear{{Conselice}}{{Conselice}}{2006}]{conselice2006}
{Conselice} C.~J.,  2006, \mn@doi [\mnras] {10.1111/j.1365-2966.2006.11114.x},
  \href {https://ui.adsabs.harvard.edu/abs/2006MNRAS.373.1389C} {373, 1389}

\bibitem[\protect\citeauthoryear{{Conselice}}{{Conselice}}{2014}]{Conselice2014}
{Conselice} C.~J.,  2014, \mn@doi [\araa]
  {10.1146/annurev-astro-081913-040037}, \href
  {https://ui.adsabs.harvard.edu/abs/2014ARA&A..52..291C} {52, 291}

\bibitem[\protect\citeauthoryear{{Conselice}, {Blackburne}  \&
  {Papovich}}{{Conselice} et~al.}{2005}]{Conselice2005}
{Conselice} C.~J.,  {Blackburne} J.~A.,   {Papovich} C.,  2005, \mn@doi [\apj]
  {10.1086/426102}, \href
  {https://ui.adsabs.harvard.edu/abs/2005ApJ...620..564C} {620, 564}

\bibitem[\protect\citeauthoryear{{DES Collaboration}}{{DES
  Collaboration}}{2005}]{DES2005}
{DES Collaboration} 2005, arXiv e-prints, \href
  {https://ui.adsabs.harvard.edu/abs/2005astro.ph.10346T} {pp
  astro--ph/0510346}

\bibitem[\protect\citeauthoryear{{DES Collaboration} et~al.,}{{DES
  Collaboration} et~al.}{2016}]{DES2016}
{DES Collaboration} et~al., 2016, \mn@doi [\mnras] {10.1093/mnras/stw641},
  \href {https://ui.adsabs.harvard.edu/abs/2016MNRAS.460.1270D} {460, 1270}

\bibitem[\protect\citeauthoryear{{De Vicente}, {S{\'a}nchez}  \&
  {Sevilla-Noarbe}}{{De Vicente} et~al.}{2016}]{DeVicente2016}
{De Vicente} J.,  {S{\'a}nchez} E.,   {Sevilla-Noarbe} I.,  2016, \mn@doi
  [\mnras] {10.1093/mnras/stw857}, \href
  {https://ui.adsabs.harvard.edu/abs/2016MNRAS.459.3078D} {459, 3078}

\bibitem[\protect\citeauthoryear{{Deeley} et~al.,}{{Deeley}
  et~al.}{2020}]{Deeley2020}
{Deeley} S.,  et~al., 2020, \mn@doi [\mnras] {10.1093/mnras/staa2417}, \href
  {https://ui.adsabs.harvard.edu/abs/2020MNRAS.498.2372D} {498, 2372}

\bibitem[\protect\citeauthoryear{{Deeley}, {Drinkwater}, {Sweet}, {Bekki},
  {Couch}, {Forbes}  \& {Dolfi}}{{Deeley} et~al.}{2021}]{Deeley2021}
{Deeley} S.,  {Drinkwater} M.~J.,  {Sweet} S.~M.,  {Bekki} K.,  {Couch} W.~J.,
  {Forbes} D.~A.,   {Dolfi} A.,  2021, \mn@doi [\mnras]
  {10.1093/mnras/stab2007}, \href
  {https://ui.adsabs.harvard.edu/abs/2021MNRAS.508..895D} {508, 895}

\bibitem[\protect\citeauthoryear{{Dom{\'\i}nguez S{\'a}nchez},
  {Huertas-Company}, {Bernardi}, {Tuccillo}  \& {Fischer}}{{Dom{\'\i}nguez
  S{\'a}nchez} et~al.}{2018}]{Dominguez-Sanchez2018}
{Dom{\'\i}nguez S{\'a}nchez} H.,  {Huertas-Company} M.,  {Bernardi} M.,
  {Tuccillo} D.,   {Fischer} J.~L.,  2018, \mn@doi [\mnras]
  {10.1093/mnras/sty338}, \href
  {https://ui.adsabs.harvard.edu/abs/2018MNRAS.476.3661D} {476, 3661}

\bibitem[\protect\citeauthoryear{{Dom{\'\i}nguez S{\'a}nchez}, {Margalef},
  {Bernardi}  \& {Huertas-Company}}{{Dom{\'\i}nguez S{\'a}nchez}
  et~al.}{2022}]{DS2022}
{Dom{\'\i}nguez S{\'a}nchez} H.,  {Margalef} B.,  {Bernardi} M.,
  {Huertas-Company} M.,  2022, \mn@doi [\mnras] {10.1093/mnras/stab3089}, \href
  {https://ui.adsabs.harvard.edu/abs/2022MNRAS.509.4024D} {509, 4024}

\bibitem[\protect\citeauthoryear{{Dressler}}{{Dressler}}{1980}]{Dressler1980}
{Dressler} A.,  1980, \mn@doi [\apj] {10.1086/157753}, \href
  {https://ui.adsabs.harvard.edu/abs/1980ApJ...236..351D} {236, 351}

\bibitem[\protect\citeauthoryear{{Dressler}, {Oemler}, {Butcher}  \&
  {Gunn}}{{Dressler} et~al.}{1994}]{Dressler1994}
{Dressler} A.,  {Oemler} Augustus J.,  {Butcher} H.~R.,   {Gunn} J.~E.,  1994,
  \mn@doi [\apj] {10.1086/174386}, \href
  {https://ui.adsabs.harvard.edu/abs/1994ApJ...430..107D} {430, 107}

\bibitem[\protect\citeauthoryear{{Everett} et~al.,}{{Everett}
  et~al.}{2022}]{Everett2022}
{Everett} S.,  et~al., 2022, \mn@doi [\apjs] {10.3847/1538-4365/ac26c1}, \href
  {https://ui.adsabs.harvard.edu/abs/2022ApJS..258...15E} {258, 15}

\bibitem[\protect\citeauthoryear{{Ferreira}, {Conselice}, {Duncan}, {Cheng},
  {Griffiths}  \& {Whitney}}{{Ferreira} et~al.}{2020}]{Ferreira2020}
{Ferreira} L.,  {Conselice} C.~J.,  {Duncan} K.,  {Cheng} T.-Y.,  {Griffiths}
  A.,   {Whitney} A.,  2020, \mn@doi [\apj] {10.3847/1538-4357/ab8f9b}, \href
  {https://ui.adsabs.harvard.edu/abs/2020ApJ...895..115F} {895, 115}

\bibitem[\protect\citeauthoryear{{Fischer}, {Dom{\'\i}nguez S{\'a}nchez}  \&
  {Bernardi}}{{Fischer} et~al.}{2019}]{Fischer2019}
{Fischer} J.~L.,  {Dom{\'\i}nguez S{\'a}nchez} H.,   {Bernardi} M.,  2019,
  \mn@doi [\mnras] {10.1093/mnras/sty3135}, \href
  {https://ui.adsabs.harvard.edu/abs/2019MNRAS.483.2057F} {483, 2057}

\bibitem[\protect\citeauthoryear{{Flaugher} et~al.,}{{Flaugher}
  et~al.}{2015}]{Flaugher2015}
{Flaugher} B.,  et~al., 2015, \mn@doi [\aj] {10.1088/0004-6256/150/5/150},
  \href {https://ui.adsabs.harvard.edu/abs/2015AJ....150..150F} {150, 150}

\bibitem[\protect\citeauthoryear{{Fukugita} et~al.,}{{Fukugita}
  et~al.}{2007}]{Fukugita2007}
{Fukugita} M.,  et~al., 2007, \mn@doi [\aj] {10.1086/518962}, \href
  {https://ui.adsabs.harvard.edu/abs/2007AJ....134..579F} {134, 579}

\bibitem[\protect\citeauthoryear{{Ghosh}, {Urry}, {Wang}, {Schawinski}, {Turp}
  \& {Powell}}{{Ghosh} et~al.}{2020}]{Ghosh2020}
{Ghosh} A.,  {Urry} C.~M.,  {Wang} Z.,  {Schawinski} K.,  {Turp} D.,   {Powell}
  M.~C.,  2020, \mn@doi [\apj] {10.3847/1538-4357/ab8a47}, \href
  {https://ui.adsabs.harvard.edu/abs/2020ApJ...895..112G} {895, 112}

\bibitem[\protect\citeauthoryear{{Gupta}, {Srijith}  \& {Desai}}{{Gupta}
  et~al.}{2022}]{Gupta2022}
{Gupta} R.,  {Srijith} P.~K.,   {Desai} S.,  2022, \mn@doi [Astronomy and
  Computing] {10.1016/j.ascom.2021.100543}, \href
  {https://ui.adsabs.harvard.edu/abs/2022A&C....3800543G} {38, 100543}

\bibitem[\protect\citeauthoryear{{Hausen} \& {Robertson}}{{Hausen} \&
  {Robertson}}{2020}]{Hausen2020}
{Hausen} R.,  {Robertson} B.~E.,  2020, \mn@doi [\apjs]
  {10.3847/1538-4365/ab8868}, \href
  {https://ui.adsabs.harvard.edu/abs/2020ApJS..248...20H} {248, 20}

\bibitem[\protect\citeauthoryear{{Holmberg}}{{Holmberg}}{1958}]{Holmberg1958}
{Holmberg} E.,  1958, Meddelanden fran Lunds Astronomiska Observatorium Serie
  II, \href {https://ui.adsabs.harvard.edu/abs/1958MeLuS.136....1H} {136, 1}

\bibitem[\protect\citeauthoryear{{Huertas-Company}, {Rouan}, {Tasca}, {Soucail}
   \& {Le F{\`e}vre}}{{Huertas-Company} et~al.}{2008}]{Huertas-Company2008}
{Huertas-Company} M.,  {Rouan} D.,  {Tasca} L.,  {Soucail} G.,   {Le F{\`e}vre}
  O.,  2008, \mn@doi [\aap] {10.1051/0004-6361:20078625}, \href
  {https://ui.adsabs.harvard.edu/abs/2008A&A...478..971H} {478, 971}

\bibitem[\protect\citeauthoryear{{Huertas-Company} et~al.,}{{Huertas-Company}
  et~al.}{2016}]{Huertas-Company2016}
{Huertas-Company} M.,  et~al., 2016, \mn@doi [\mnras] {10.1093/mnras/stw1866},
  \href {https://ui.adsabs.harvard.edu/abs/2016MNRAS.462.4495H} {462, 4495}

\bibitem[\protect\citeauthoryear{{Huertas-Company} et~al.,}{{Huertas-Company}
  et~al.}{2018}]{Huertas-Company2018}
{Huertas-Company} M.,  et~al., 2018, \mn@doi [\apj] {10.3847/1538-4357/aabfed},
  \href {https://ui.adsabs.harvard.edu/abs/2018ApJ...858..114H} {858, 114}

\bibitem[\protect\citeauthoryear{{Ivezi{\'c}} et~al.,}{{Ivezi{\'c}}
  et~al.}{2019}]{Ivezic2019}
{Ivezi{\'c}} {\v{Z}}.,  et~al., 2019, \mn@doi [\apj]
  {10.3847/1538-4357/ab042c}, \href
  {https://ui.adsabs.harvard.edu/abs/2019ApJ...873..111I} {873, 111}

\bibitem[\protect\citeauthoryear{{Jacobs}, {Glazebrook}, {Collett}, {More}  \&
  {McCarthy}}{{Jacobs} et~al.}{2017}]{Jacobs2017}
{Jacobs} C.,  {Glazebrook} K.,  {Collett} T.,  {More} A.,   {McCarthy} C.,
  2017, \mn@doi [\mnras] {10.1093/mnras/stx1492}, \href
  {https://ui.adsabs.harvard.edu/abs/2017MNRAS.471..167J} {471, 167}

\bibitem[\protect\citeauthoryear{{Krywult} et~al.,}{{Krywult}
  et~al.}{2017}]{Krywult2017}
{Krywult} J.,  et~al., 2017, \mn@doi [\aap] {10.1051/0004-6361/201628953},
  \href {https://ui.adsabs.harvard.edu/abs/2017A&A...598A.120K} {598, A120}

\bibitem[\protect\citeauthoryear{{Lahav} et~al.,}{{Lahav}
  et~al.}{1995}]{Lahav1995}
{Lahav} O.,  et~al., 1995, \mn@doi [Science] {10.1126/science.267.5199.859},
  \href {https://ui.adsabs.harvard.edu/abs/1995Sci...267..859L} {267, 859}

\bibitem[\protect\citeauthoryear{{Lanusse}, {Ma}, {Li}, {Collett}, {Li},
  {Ravanbakhsh}, {Mandelbaum}  \& {P{\'o}czos}}{{Lanusse}
  et~al.}{2018}]{Lanusse2018}
{Lanusse} F.,  {Ma} Q.,  {Li} N.,  {Collett} T.~E.,  {Li} C.-L.,  {Ravanbakhsh}
  S.,  {Mandelbaum} R.,   {P{\'o}czos} B.,  2018, \mn@doi [\mnras]
  {10.1093/mnras/stx1665}, \href
  {https://ui.adsabs.harvard.edu/abs/2018MNRAS.473.3895L} {473, 3895}

\bibitem[\protect\citeauthoryear{{Laureijs} et~al.,}{{Laureijs}
  et~al.}{2011}]{laureijs2011}
{Laureijs} R.,  et~al., 2011, arXiv e-prints, \href
  {https://ui.adsabs.harvard.edu/abs/2011arXiv1110.3193L} {p. arXiv:1110.3193}

\bibitem[\protect\citeauthoryear{{Lintott} et~al.,}{{Lintott}
  et~al.}{2008}]{Lintott2008}
{Lintott} C.~J.,  et~al., 2008, \mn@doi [\mnras]
  {10.1111/j.1365-2966.2008.13689.x}, \href
  {https://ui.adsabs.harvard.edu/abs/2008MNRAS.389.1179L} {389, 1179}

\bibitem[\protect\citeauthoryear{{Lintott} et~al.,}{{Lintott}
  et~al.}{2011}]{Lintott2011}
{Lintott} C.,  et~al., 2011, \mn@doi [\mnras]
  {10.1111/j.1365-2966.2010.17432.x}, \href
  {https://ui.adsabs.harvard.edu/abs/2011MNRAS.410..166L} {410, 166}

\bibitem[\protect\citeauthoryear{{Martel}, {Premadi}  \& {Matzner}}{{Martel}
  et~al.}{1998}]{Martel1998}
{Martel} H.,  {Premadi} P.,   {Matzner} R.,  1998, \mn@doi [\apj]
  {10.1086/305472}, \href
  {https://ui.adsabs.harvard.edu/abs/1998ApJ...497..512M} {497, 512}

\bibitem[\protect\citeauthoryear{{Moutard} et~al.,}{{Moutard}
  et~al.}{2016a}]{Moutard2016a}
{Moutard} T.,  et~al., 2016a, \mn@doi [\aap] {10.1051/0004-6361/201527945},
  \href {https://ui.adsabs.harvard.edu/abs/2016A&A...590A.102M} {590, A102}

\bibitem[\protect\citeauthoryear{{Moutard} et~al.,}{{Moutard}
  et~al.}{2016b}]{Moutard2016b}
{Moutard} T.,  et~al., 2016b, \mn@doi [\aap] {10.1051/0004-6361/201527294},
  \href {https://ui.adsabs.harvard.edu/abs/2016A&A...590A.103M} {590, A103}

\bibitem[\protect\citeauthoryear{{Nair} \& {Abraham}}{{Nair} \&
  {Abraham}}{2010}]{Nair2010}
{Nair} P.~B.,  {Abraham} R.~G.,  2010, \mn@doi [\apjs]
  {10.1088/0067-0049/186/2/427}, \href
  {https://ui.adsabs.harvard.edu/abs/2010ApJS..186..427N} {186, 427}

\bibitem[\protect\citeauthoryear{{Neilsen}, {Annis}, {Diehl}, {Swanson},
  {D'Andrea}, {Kent}  \& {Drlica-Wagner}}{{Neilsen} et~al.}{2019}]{Neilsen2019}
{Neilsen} Eric~H. J.,  {Annis} J.~T.,  {Diehl} H.~T.,  {Swanson} M. E.~C.,
  {D'Andrea} C.,  {Kent} S.,   {Drlica-Wagner} A.,  2019, arXiv e-prints, \href
  {https://ui.adsabs.harvard.edu/abs/2019arXiv191206254N} {p. arXiv:1912.06254}

\bibitem[\protect\citeauthoryear{{Odewahn}, {Stockwell}, {Pennington},
  {Humphreys}  \& {Zumach}}{{Odewahn} et~al.}{1992}]{Odewahn1992}
{Odewahn} S.~C.,  {Stockwell} E.~B.,  {Pennington} R.~L.,  {Humphreys} R.~M.,
  {Zumach} W.~A.,  1992, \mn@doi [\aj] {10.1086/116063}, \href
  {https://ui.adsabs.harvard.edu/abs/1992AJ....103..318O} {103, 318}

\bibitem[\protect\citeauthoryear{{Palmese} et~al.,}{{Palmese}
  et~al.}{2020}]{Palmese2020}
{Palmese} A.,  et~al., 2020, \mn@doi [\mnras] {10.1093/mnras/staa526}, \href
  {https://ui.adsabs.harvard.edu/abs/2020MNRAS.493.4591P} {493, 4591}

\bibitem[\protect\citeauthoryear{{Petrillo} et~al.,}{{Petrillo}
  et~al.}{2017}]{Petrillo2017}
{Petrillo} C.~E.,  et~al., 2017, \mn@doi [\mnras] {10.1093/mnras/stx2052},
  \href {https://ui.adsabs.harvard.edu/abs/2017MNRAS.472.1129P} {472, 1129}

\bibitem[\protect\citeauthoryear{{Pozzetti} et~al.,}{{Pozzetti}
  et~al.}{2010}]{Pozzetti2010}
{Pozzetti} L.,  et~al., 2010, \mn@doi [\aap] {10.1051/0004-6361/200913020},
  \href {https://ui.adsabs.harvard.edu/abs/2010A&A...523A..13P} {523, A13}

\bibitem[\protect\citeauthoryear{{Sandage}}{{Sandage}}{1961}]{Sandage1961}
{Sandage} A.,  1961, {The Hubble Atlas of Galaxies}

\bibitem[\protect\citeauthoryear{{Schuldt}, {Suyu}, {Ca{\~n}ameras},
  {Taubenberger}, {Meinhardt}, {Leal-Taix{\'e}}  \& {Hsieh}}{{Schuldt}
  et~al.}{2021}]{Schuldt2021}
{Schuldt} S.,  {Suyu} S.~H.,  {Ca{\~n}ameras} R.,  {Taubenberger} S.,
  {Meinhardt} T.,  {Leal-Taix{\'e}} L.,   {Hsieh} B.~C.,  2021, \mn@doi [\aap]
  {10.1051/0004-6361/202039945}, \href
  {https://ui.adsabs.harvard.edu/abs/2021A&A...651A..55S} {651, A55}

\bibitem[\protect\citeauthoryear{{Scodeggio} et~al.,}{{Scodeggio}
  et~al.}{2018}]{Scodeggio2018}
{Scodeggio} M.,  et~al., 2018, \mn@doi [\aap] {10.1051/0004-6361/201630114},
  \href {https://ui.adsabs.harvard.edu/abs/2018A&A...609A..84S} {609, A84}

\bibitem[\protect\citeauthoryear{{Sevilla-Noarbe} et~al.,}{{Sevilla-Noarbe}
  et~al.}{2021}]{Sevilla-Noarbe2021}
{Sevilla-Noarbe} I.,  et~al., 2021, \mn@doi [\apjs] {10.3847/1538-4365/abeb66},
  \href {https://ui.adsabs.harvard.edu/abs/2021ApJS..254...24S} {254, 24}

\bibitem[\protect\citeauthoryear{{Siudek} et~al.,}{{Siudek}
  et~al.}{2018a}]{Siudek2018a}
{Siudek} M.,  et~al., 2018a, arXiv e-prints, \href
  {https://ui.adsabs.harvard.edu/abs/2018arXiv180509905S} {p. arXiv:1805.09905}

\bibitem[\protect\citeauthoryear{{Siudek} et~al.,}{{Siudek}
  et~al.}{2018b}]{Siudek2018b}
{Siudek} M.,  et~al., 2018b, \mn@doi [\aap] {10.1051/0004-6361/201832784},
  \href {https://ui.adsabs.harvard.edu/abs/2018A&A...617A..70S} {617, A70}

\bibitem[\protect\citeauthoryear{{Soo} et~al.,}{{Soo} et~al.}{2021}]{Soo2021}
{Soo} J. Y.~H.,  et~al., 2021, \mn@doi [\mnras] {10.1093/mnras/stab711}, \href
  {https://ui.adsabs.harvard.edu/abs/2021MNRAS.503.4118S} {503, 4118}

\bibitem[\protect\citeauthoryear{{Soumagnac} et~al.,}{{Soumagnac}
  et~al.}{2015}]{Soumagnac2015}
{Soumagnac} M.~T.,  et~al., 2015, \mn@doi [\mnras] {10.1093/mnras/stu1410},
  \href {https://ui.adsabs.harvard.edu/abs/2015MNRAS.450..666S} {450, 666}

\bibitem[\protect\citeauthoryear{{Tarsitano} et~al.,}{{Tarsitano}
  et~al.}{2018}]{Tarsitano2018}
{Tarsitano} F.,  et~al., 2018, \mn@doi [\mnras] {10.1093/mnras/sty1970}, \href
  {https://ui.adsabs.harvard.edu/abs/2018MNRAS.481.2018T} {481, 2018}

\bibitem[\protect\citeauthoryear{{Tasca} et~al.,}{{Tasca}
  et~al.}{2009}]{Tasca2009}
{Tasca} L.~A.~M.,  et~al., 2009, \mn@doi [\aap] {10.1051/0004-6361/200912213},
  \href {https://ui.adsabs.harvard.edu/abs/2009A&A...503..379T} {503, 379}

\bibitem[\protect\citeauthoryear{{Tohill}, {Ferreira}, {Conselice}, {Bamford}
  \& {Ferrari}}{{Tohill} et~al.}{2021}]{Tohill2021}
{Tohill} C.,  {Ferreira} L.,  {Conselice} C.~J.,  {Bamford} S.~P.,   {Ferrari}
  F.,  2021, \mn@doi [\apj] {10.3847/1538-4357/ac033c}, \href
  {https://ui.adsabs.harvard.edu/abs/2021ApJ...916....4T} {916, 4}

\bibitem[\protect\citeauthoryear{{Turner} et~al.,}{{Turner}
  et~al.}{2021}]{Turner2021}
{Turner} S.,  et~al., 2021, \mn@doi [\mnras] {10.1093/mnras/stab653}, \href
  {https://ui.adsabs.harvard.edu/abs/2021MNRAS.503.3010T} {503, 3010}

\bibitem[\protect\citeauthoryear{{Vega-Ferrero} et~al.,}{{Vega-Ferrero}
  et~al.}{2021}]{Vega-Ferrero2021}
{Vega-Ferrero} J.,  et~al., 2021, \mn@doi [\mnras] {10.1093/mnras/stab594},
  \href {https://ui.adsabs.harvard.edu/abs/2021MNRAS.506.1927V} {506, 1927}

\bibitem[\protect\citeauthoryear{{Walmsley} et~al.,}{{Walmsley}
  et~al.}{2020}]{Walmsley2020}
{Walmsley} M.,  et~al., 2020, \mn@doi [\mnras] {10.1093/mnras/stz2816}, \href
  {https://ui.adsabs.harvard.edu/abs/2020MNRAS.491.1554W} {491, 1554}

\bibitem[\protect\citeauthoryear{{Weir}, {Fayyad}  \& {Djorgovski}}{{Weir}
  et~al.}{1995}]{Weir1995}
{Weir} N.,  {Fayyad} U.~M.,   {Djorgovski} S.,  1995, \mn@doi [\aj]
  {10.1086/117459}, \href
  {https://ui.adsabs.harvard.edu/abs/1995AJ....109.2401W} {109, 2401}

\bibitem[\protect\citeauthoryear{{Willett} et~al.,}{{Willett}
  et~al.}{2013}]{Willett2013}
{Willett} K.~W.,  et~al., 2013, \mn@doi [\mnras] {10.1093/mnras/stt1458}, \href
  {https://ui.adsabs.harvard.edu/abs/2013MNRAS.435.2835W} {435, 2835}

\bibitem[\protect\citeauthoryear{{Wuyts} et~al.,}{{Wuyts}
  et~al.}{2011}]{Wuyts2011}
{Wuyts} S.,  et~al., 2011, \mn@doi [\apj] {10.1088/0004-637X/742/2/96}, \href
  {https://ui.adsabs.harvard.edu/abs/2011ApJ...742...96W} {742, 96}

\bibitem[\protect\citeauthoryear{{de Vaucouleurs}}{{de
  Vaucouleurs}}{1959}]{deVaucouleurs1959}
{de Vaucouleurs} G.,  1959, \mn@doi [Handbuch der Physik]
  {10.1007/978-3-642-45932-0_7}, \href
  {https://ui.adsabs.harvard.edu/abs/1959HDP....53..275D} {53, 275}

\bibitem[\protect\citeauthoryear{{de Vaucouleurs}}{{de
  Vaucouleurs}}{1964}]{deVaucouleurs1964}
{de Vaucouleurs} G.,  1964, \mn@doi [\aj] {10.1086/109329}, \href
  {https://ui.adsabs.harvard.edu/abs/1964AJ.....69..561D} {69, 561}

\makeatother
\end{thebibliography}

\appendix
\section*{Affiliations}
$^{1}$ Centre for Extragalactic Astronomy, Durham University, South Road, Durham DH1 3LE, UK\\
$^{2}$ Institute of Space Sciences (ICE, CSIC), Campus UAB, Carrerde Can Magrans, s/n, 08193 Barcelona, Spain\\
$^{3}$ Instituto de Astrof\'{\i}ısica de Canarias (IAC) La Laguna, 38205, Spain\\
$^{4}$ Jodrell Bank Centre for Astrophysics, University of Manchester, Oxford Road, Manchester M13 9PL, UK\\
$^{5}$ Institut de F\'{\i}sica d'Altes Energies (IFAE), The Barcelona Institute of Science and Technology, 08193 Bellaterra (Barcelona), Spain \\
$^{6}$ School of Physics and Astronomy, University of Nottingham, University Park, Nottingham, NG7 2RD, UK\\
$^{7}$ Department of Physics and Astronomy, University of Pennsylvania, Philadelphia, PA 19104, USA\\
$^{8}$ LERMA, Observatoire de Paris, CNRS, PSL, Universit\'e Paris Diderot, France\\
$^{9}$ Institute of Physics, Jan Kochanowski University, ul. Uniwersytecka 7, 25-406 Kielce, Poland\\
$^{10}$ Department of Physics, University of California Berkeley, 366 LeConte Hall MC 7300, Berkeley, CA, 94720, USA, NASA Einstein fellow\\
$^{11}$ Laborat\'orio Interinstitucional de e-Astronomia - LIneA, Rua Gal. Jos\'e Cristino 77, Rio de Janeiro, RJ - 20921-400, Brazil\\
$^{12}$ Observat\'orio Nacional, Rua Gal. Jos\'e Cristino 77, Rio de Janeiro, RJ - 20921-400, Brazil\\
$^{13}$ Department of Astrophysical Sciences, Princeton University, Peyton Hall, Princeton, NJ 08544, USA\\
$^{14}$ Universidad de La Laguna, Dpto. Astrofísica, E-38206 La Laguna, Tenerife, Spain\\
$^{15}$ University Observatory, Faculty of Physics, Ludwig-Maximilians-Universit\"at, Scheinerstr. 1, 81679 Munich, Germany\\
$^{16}$ Institute of Cosmology and Gravitation, University of Portsmouth, Portsmouth, PO1 3FX, UK\\
$^{17}$ Department of Physics \& Astronomy, University College London, Gower Street, London, WC1E 6BT, UK\\
$^{18}$ Center for Astrophysics $\vert$ Harvard \& Smithsonian, 60 Garden Street, Cambridge, MA 02138, USA\\
$^{19}$ Santa Cruz Institute for Particle Physics, Santa Cruz, CA 95064, USA\\
$^{20}$ Center for Astrophysical Surveys, National Center for Supercomputing Applications, 1205 West Clark St., Urbana, IL 61801, USA\\
$^{21}$ Computer Science and Mathematics Division, Oak Ridge National Laboratory, Oak Ridge, TN 37831\\
$^{22}$ Centro de Investigaciones Energ\'eticas, Medioambientales y Tecnol\'ogicas (CIEMAT), Madrid, Spain\\
$^{23}$ Department of Astronomy, University of Illinois at Urbana-Champaign, 1002 W. Green Street, Urbana, IL 61801, USA\\
$^{24}$ Institute of Astronomy, University of Cambridge, Madingley Road, Cambridge CB3 0HA, UK\\
$^{25}$ Fermi National Accelerator Laboratory, P. O. Box 500, Batavia, IL 60510, USA\\
$^{26}$ Department of Physics, University of Michigan, Ann Arbor, MI 48109, USA\\
$^{27}$ Institute of Theoretical Astrophysics, University of Oslo. P.O. Box 1029 Blindern, NO-0315 Oslo, Norway\\
$^{28}$ Kavli Institute for Cosmological Physics, University of Chicago, Chicago, IL 60637, USA\\
$^{29}$ Instituto de Fisica Teorica UAM/CSIC, Universidad Autonoma de Madrid, 28049 Madrid, Spain\\
$^{30}$ Center for Cosmology and Astro-Particle Physics, The Ohio State University, Columbus, OH 43210, USA\\
$^{31}$ Department of Physics, The Ohio State University, Columbus, OH 43210, USA\\
$^{32}$ Australian Astronomical Optics, Macquarie University, North Ryde, NSW 2113, Australia\\
$^{33}$ Lowell Observatory, 1400 Mars Hill Rd, Flagstaff, AZ 86001, USA\\
$^{34}$ Hamburger Sternwarte, Universit\"{a}t Hamburg, Gojenbergsweg 112, 21029 Hamburg, Germany\\
$^{35}$ School of Physics and Astronomy, University of Southampton,  Southampton, SO17 1BJ, UK\\
$^{36}$ Lawrence Berkeley National Laboratory, 1 Cyclotron Road, Berkeley, CA 94720, USA\\
$^{37}$ Instituci\'o Catalana de Recerca i Estudis Avan\c{c}ats, E-08010 Barcelona, Spain\\
$^{38}$ School of Mathematics and Physics, University of Queensland,  Brisbane, QLD 4072, Australia\\
$^{39}$ Department of Physics, Carnegie Mellon University, Pittsburgh, Pennsylvania 15312, USA\\
$^{40}$ NSF AI Planning Institute for Physics of the Future, Carnegie Mellon University, Pittsburgh, PA 15213, USA\\
$^{41}$ Department of Physics, IIT Hyderabad, Kandi, Telangana 502285, India\\
$^{42}$ Jet Propulsion Laboratory, California Institute of Technology, 4800 Oak Grove Dr., Pasadena, CA 91109, USA\\

\section{Figures of Section~4}
\begin{figure*}{}
\begin{center}
\graphicspath{}
	\includegraphics[width=2.1\columnwidth]{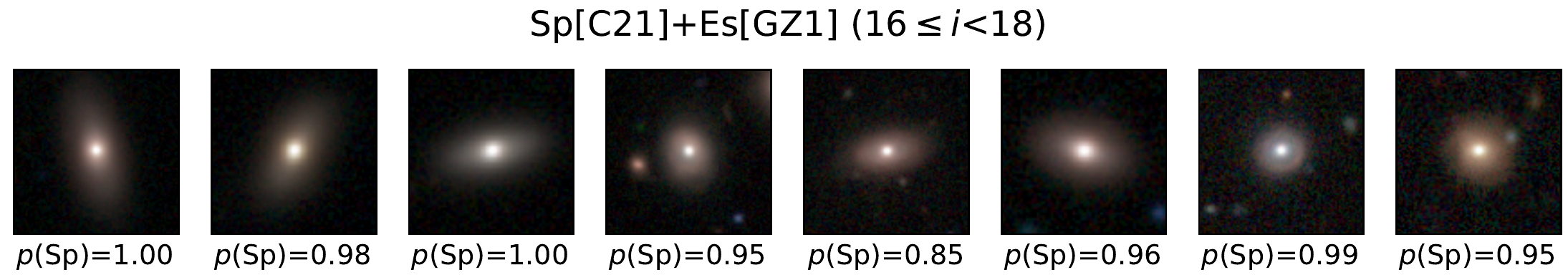}
   	\caption{Randomly selected examples of the mismatches that are classified as Sp by C21 and labelled as Es in GZ1. The $p$(Sp) values quoted under each panel is the probability of being Sp provided by C21.}
    \label{fig:example_S+GZ1_Sp+Es}
\end{center}
\end{figure*}

\begin{figure*}{}
\begin{center}
\graphicspath{}
	\includegraphics[width=2.1\columnwidth]{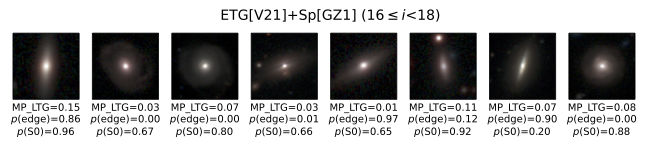}
   	\caption{Randomly selected examples of the mismatches that are classified as ETG by V21 and labelled as spiral galaxies in GZ1. The MP\_LTG, $p$(edge), and $p$(S0) values quoted under each panel represents the median probability of being LTG, edge-on galaxies (provided by V21), and lenticular galaxies (S0; provided by DS18), respectively. Note that most of these are actually edge-on and/or lenticular galaxies.}
    \label{fig:example_V+GZ1_ETG+Sp}
\end{center}
\end{figure*}

\begin{figure*}{}
\begin{center}
\graphicspath{}
	\includegraphics[width=2.1\columnwidth]{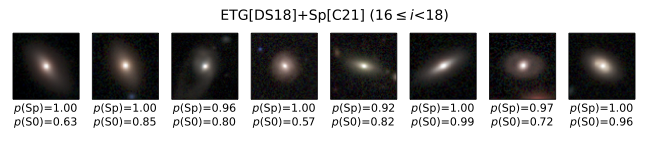}
   	\caption{Randomly selected examples of the mismatches that are classified as spiral galaxies (Sp) by C21 and labelled as ETG in DS18. The $p$(Sp) is the probability of being a spiral galaxy according to C21 while the $p$(S0) represents the probability of being a lenticular galaxy, according to DS18.}
    \label{fig:example_S+DS18_Sp+ETG}
\end{center}
\end{figure*}

\label{lastpage}
\end{document}